\documentstyle{mn}
\input{psfig}
\def\simgt{\hbox{\rlap{\raise 0.425ex\hbox{$>$}}\lower 0.65ex\hbox{$\sim$}}}
\def\simlt{\hbox{\rlap{\raise 0.425ex\hbox{$<$}}\lower 0.65ex\hbox{$\sim$}}}

\def\degree{^\circ}

\def\bj {b_{\rm J}}
\def\lb {L_{\rm b_{\rm J}}}
\def\mb {M_{\rm b_{\rm J}}}
\def\kms {km$\,$s$^{-1}$}

\def\kmsmpc {km$\,$s$^{-1}$Mpc$^{-1}$}
\def\om {\Omega_{\rm m}}
\def\ol {\Omega_{\Lambda}}
\def \ho {H_0}

\def \mgii {Mg{\small~II}}
\def \caii {Ca{\small~II}}

\def \cah {Ca{\small~II}~H}
\def \civ {C{\small~IV}}
\def \ciii {C{\small~III}]}

\def \fs {f_{\rm s}}
\def \fc {f_{\rm c}}
\def \nq {N_{\rm Q}}
\def \lp {\log\Phi}
\def \dlp {\Delta\log\Phi}
\def \nobs {N_{\rm obs}}
\def \npre {N_{\rm pred}}
\def \aj {AJ}
\def \mnras {MNRAS}
\def \apj {ApJ}

\title[The 2QZ spectroscopic catalogue]{The 2dF QSO Redshift Survey - XII.  
The spectroscopic catalogue and luminosity function}

\author[S.~M. Croom et al.]
	{S.~M. Croom$^{1}$\thanks{scroom@aaoepp.aao.gov.au},
	R.~J. Smith$^{2}$, B.~J. Boyle$^{1}$, T. Shanks$^{3}$,
	L. Miller$^{4}$, P.~J. Outram$^{3}$,
	\newauthor N.~S. Loaring$^{4,5}$\\
${^1}$ Anglo-Australian Observatory, PO Box 296, Epping, NSW 1710, 
Australia \\ 
${^2}$ Astrophysics Research Institute, Liverpool John Moores
	University, Twelve Quays House, Egerton Wharf, Birkenhead,
	CH41 1LD, UK\\
${^3}$ Department of Physics, University of Durham, South Road, 
Durham, DH1 3LE, UK \\
${^4}$ Department of Physics, Oxford University, 1 Keble Road, Oxford,
OX1 3RH, UK\\
${^5}$ Mullard Space Science Laboratory, Holmbury St. Mary, Dorking,
Surrey, RH5 6NT, UK\\
}

\begin{document}

\maketitle

\newcommand{\fmmm}[1]{\mbox{$#1$}}
\newcommand{\scnd}{\mbox{\fmmm{''}\hskip-0.3em .}}
\newcommand{\scnp}{\mbox{\fmmm{''}}}

\begin{abstract}

We present the final catalogue of the 2dF QSO Redshift Survey (2QZ),
based on Anglo-Australian Telescope 2dF spectroscopic observations of
44576 colour-selected ($u\bj r$) objects with $18.25<\bj<20.85$
selected from APM scans of UK Schmidt Telescope (UKST) photographic
plates. The 2QZ comprises 23338 QSOs, 12292 galactic stars (including
2071 white dwarfs) and 4558 compact narrow-emission-line galaxies.  We
obtained a reliable spectroscopic identification for 86 per cent of
objects observed with 2dF. We also report on the 6dF QSO Redshift
Survey (6QZ), based on UKST 6dF observations of 1564 brighter
$16<\bj<18.25$ sources selected from the same photographic input
catalogue.  In total, we identified 322 QSOs spectroscopically in the
6QZ.  The completed 2QZ is, by more than a factor 50, the largest
homogeneous QSO catalogue ever constructed at these faint limits
($\bj<20.85$) and high QSO surface densities (35 QSOs deg$^{-2}$).  As
such it represents an important resource in the study of the Universe
at moderate-to-high redshifts.  As an example of the results possible
with the 2QZ, we also present our most recent analysis of the optical
QSO luminosity function and its cosmological evolution with  redshift.
For a flat, $\om=0.3$ and $\ol=0.7$, Universe, we find that a double
power law with luminosity evolution that is exponential in look-back
time, $\tau$, of the form $\lb^*(z)\propto e^{6.15\tau}$, equivalent
to an e-folding time of 2Gyr, provides an acceptable fit to the
redshift dependence of the QSO luminosity function over the range $0.4
< z < 2.1$ and $\mb<-22.5$.  Evolution described by a quadratic in
redshift is also an acceptable fit, with
$\lb^*(z)\propto10^{1.39z-0.29z^2}$.  

\end{abstract}
\begin{keywords}
quasars: general\ -- galaxies: active\ -- galaxies: Seyfert\ -- stars:
white dwarfs\ -- catalogues\ -- surveys 
\end{keywords}
\section{Introduction}

\begin{figure*}
\centering
\centerline{\psfig{file=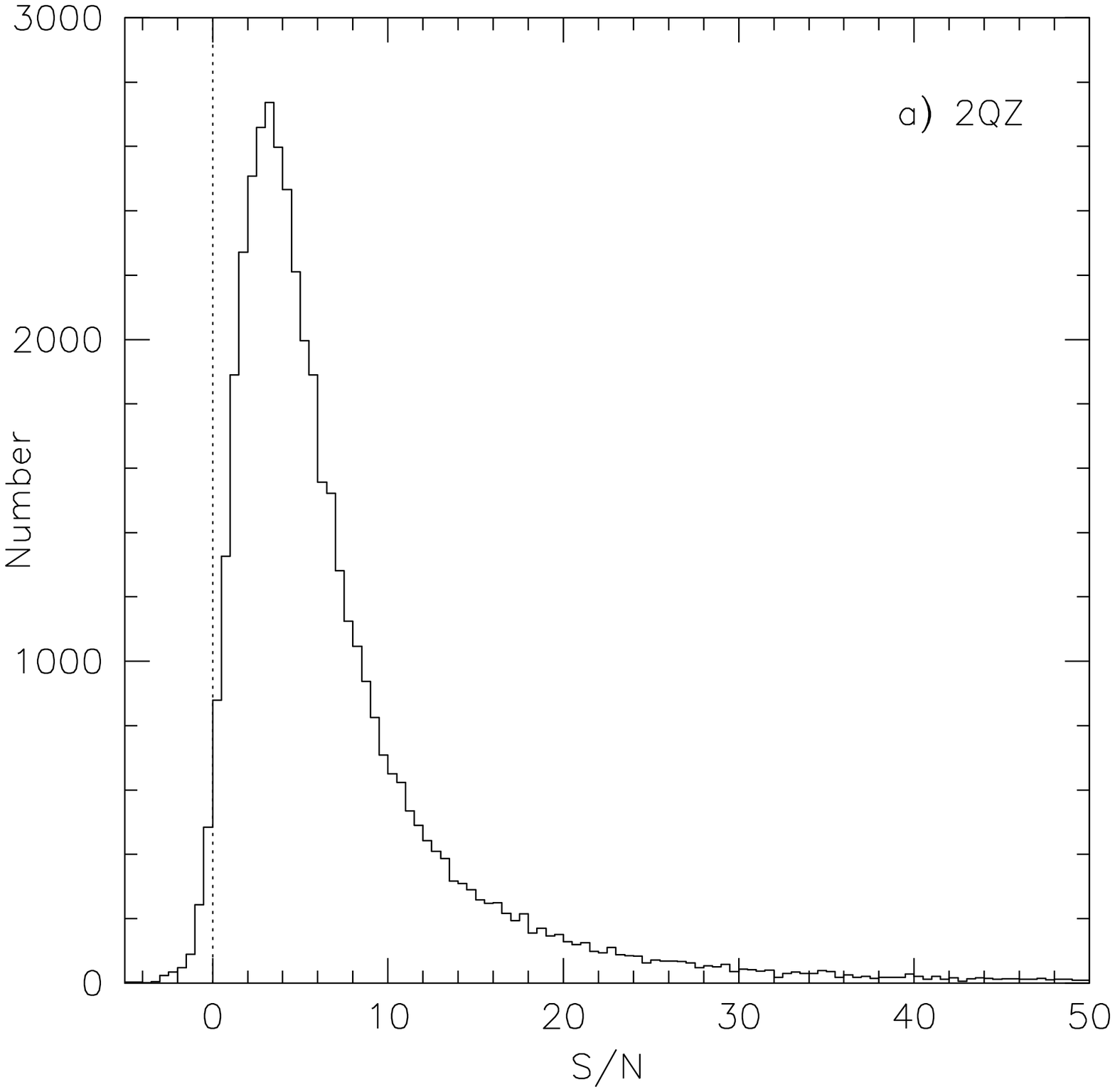,width=8cm}\psfig{file=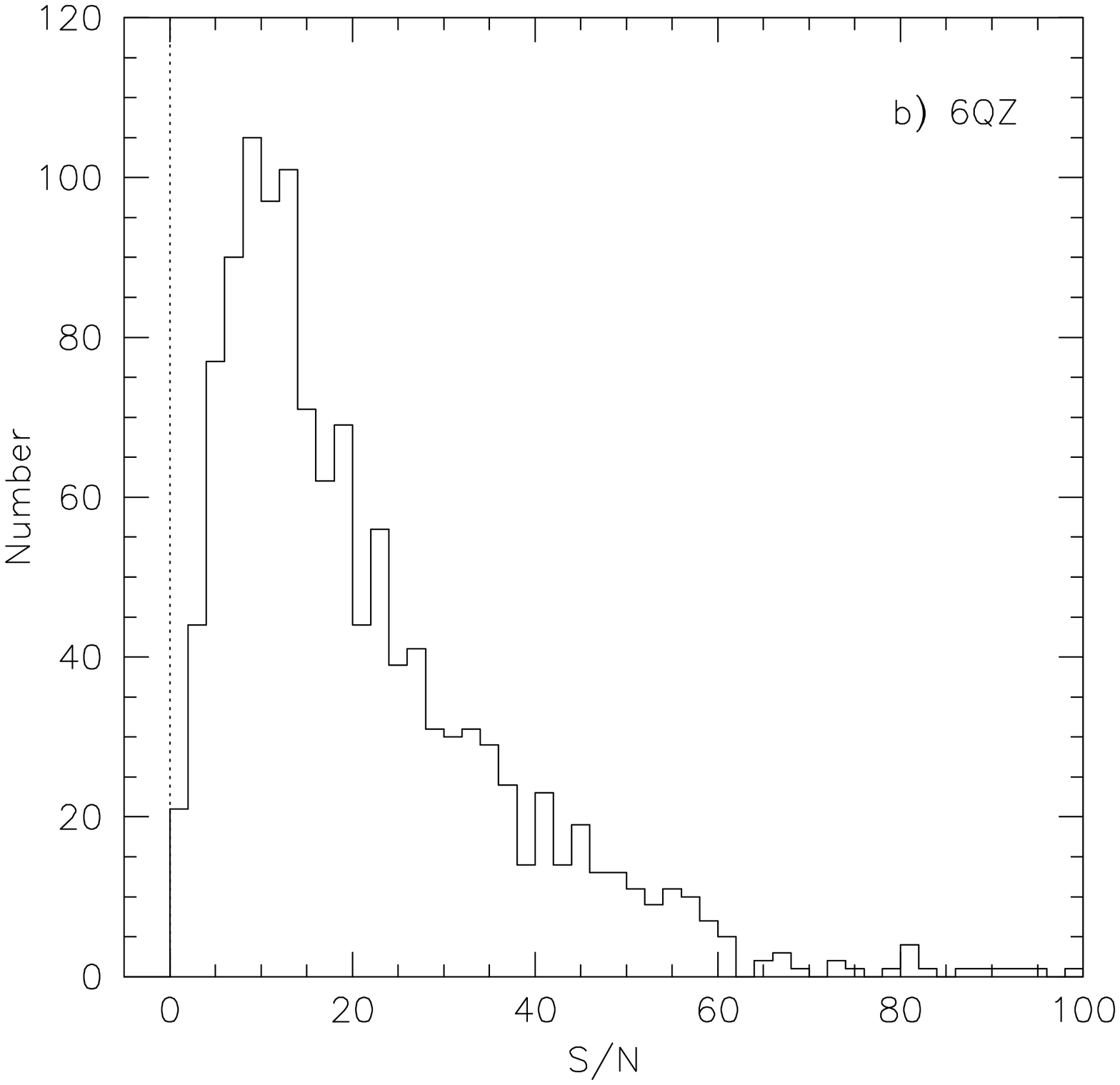,width=8cm}}
\caption{The distribution of measured spectral signal-to-noise per pixel
in the $B$-band for the best observation of each object in the  a) 2QZ and b)
6QZ catalogues.  Note that a small number of sources have $S/N<0$ due
to residual errors in sky subtraction of faint sources.}
\label{fig:sn}
\end{figure*}

The new generation of QSO surveys provide us with an unparalleled
database with which to study the properties of the QSO population.  In
this paper we report on the completion of 2dF QSO Redshift Survey
(2QZ) and the associated 6dF QSO Redshift Survey (6QZ).  These surveys
provide almost 25000 QSOs in a single homogeneous database, covering
almost five magnitudes ($16<\bj<20.85$).  When combined with QSOs from
complementary spectroscopic  observations carried out as part of the
Sloan Digital Sky Survey (SDSS; e.g. York et al. 2000), the number of
known QSOs has increased more than five-fold in  the past five years.
Furthermore, the vast majority of known QSOs are now in a few
homogeneous samples with well-defined selection criteria, rather than
a  highly heterogeneous assemblage of small surveys and serendipitous
discoveries.  The properties of the QSO population may now be
determined with a new level of statistical precision.

The 2QZ provides a valuable resource to study the large-scale
structure of the Universe on the largest scales over a wide range of
redshifts.  It can also be used to directly probe the mass
distribution via lensing studies.  In addition, the 2QZ also contains
significant new catalogues of other classes of astronomical objects,
e.g. white dwarfs \cite{vennes02} and cataclysmic variables
\cite{marsh02}.
 
The 2QZ sample presented in this paper has already provided new
information on the power spectrum of QSOs \cite{2qzpaper11}, the  QSO
correlation function \cite{2qzpaper2} and QSO spectral properties
\cite{composite02,composite03}.  It has also been used to search for
rare/unusual objects \cite{2bl02}, carry out lensing studies
\cite{2qzpaper10,miller03} and place limits on cosmological parameters
\cite{2qzpaper6,2qzpaper7}.

We also provide an updated estimate of the QSO optical luminosity
function (OLF) and its evolution with redshift based on the 2QZ sample
in this paper.  In this we include a new sample of QSOs obtained from
observations obtained with the 6dF facility on the UK Schmidt
Telescope (UKST).  The 6dF sample is based on the same input catalogue
as the 2QZ, but extends the coverage of the LF at any given redshift
to almost a factor of 100 in luminosity and over 1000 in space
density.  This will aid tests of the claimed departure from pure
luminosity evolution at higher luminosities \cite{hfc93,lc97}.

We describe the 2QZ and 6QZ surveys in Section \ref{sec:data} and then
discuss their selection effects and completeness in Section
\ref{sec:comp}.  In  Section  \ref{sec:lf} we present our analysis of
the QSO luminosity function.

\section{data}\label{sec:data}

\subsection{Input catalogue}

The selection of the QSO candidates for the 2QZ and 6QZ surveys was
based on broadband $u\bj r$ colours from APM measurements of UKST
photographic plates. Full details are given by Smith et al. (2003),
and so only a brief overview will be given here.  The survey area
comprises 30 UKST fields, arranged in two $75\degree \times 5\degree$
declination strips, one passing across the South Galactic Cap centred
on $\delta = -30\degree$ (the SGP strip) and the other across the
North Galactic Cap centred on $\delta = 0\degree$ (referred to in this
paper as the equatorial strip, but also know as the NGP strip).  The
SGP strip extends from $\alpha$ = 21$^{\rm h}$40 to $\alpha$ = 3$^{\rm
h}$15 and the equatorial strip from $\alpha$ = 9$^{\rm h}$50 to
$\alpha$ = 14$^{\rm h}$50 (B1950).  Note that the survey was
originally defined in the B1950 coordinate system, although the
publically available catalogue is presented with J2000 positions.  The
total survey area is 721.6 deg$^2$, when allowance is made for regions
of sky excised around bright stars.

In each UKST field, APM measurements of one $\bj$ plate, one $r$ plate
and up to four $u$ plates/films were used to generate a catalogue of
stellar objects with $16< \bj < 20.85$.  A sophisticated procedure was
devised to ensure catalogue homogeneity.  Corrections were made for
vignetting and field effects due to variable desensitization in the
UKST plates, these effects being particularly noticeable at the edges
of plates \cite{2qzpaper3,smith98,croom97}.  Candidates were selected
for the 2QZ based on fulfilling at least one of the following colour
criteria: $u-\bj\leq-0.36$; $u-\bj<0.12-0.8(\bj-r)$; $\bj-r<0.05$.

The $u-\bj$ limit was tightened to $u-\bj\leq-0.50$ for the 6QZ sample
($\bj\leq18.25$) to reduce the high fraction of contamination by
Galactic stars.  With these criteria, the 2QZ input catalogue
comprised 47768 candidates and the 6QZ comprised 1657 candidates
(SGP strip only, see below).

\subsection{Spectroscopic observations}

\begin{figure*}
\centering
\centerline{\psfig{file=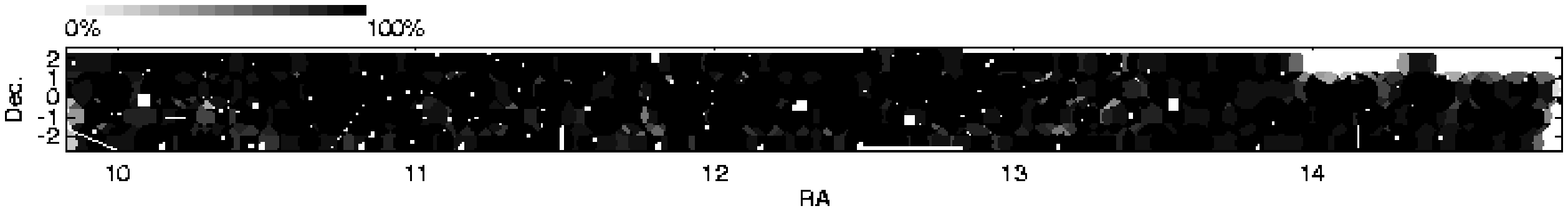,width=18cm,angle=90}}
\centerline{\psfig{file=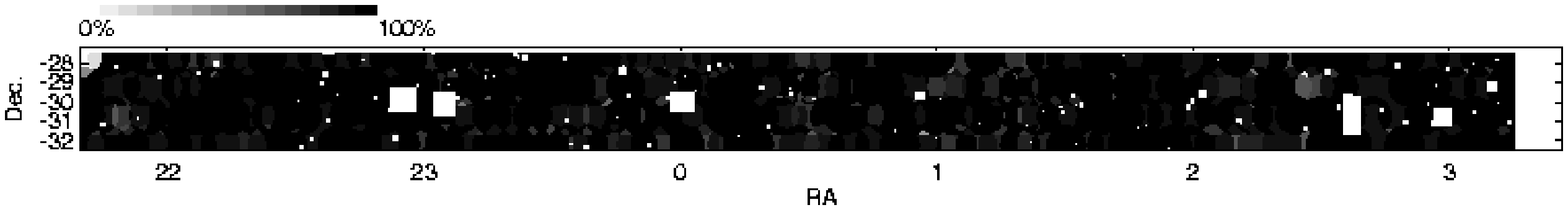,width=18cm,angle=90}}
\caption{The coverage map for the 2QZ catalogue for the equatorial
(top) and SGP (bottom) regions.  The grey-scale indicates the
percentage of 2QZ candidates observed over the two survey strips.
Regions removed due to bright stars or plate features/defects can be
seen, as can the unobserved fields at the 14$^h$ end of the
equatorial strip.}
\label{fig:cover}
\end{figure*}

Spectroscopic observations of the input catalogue were made with the
2-degree Field (2dF) instrument at the Anglo-Australian Telescope (the
2QZ sample)  and the 6-degree Field (6dF) instrument at the UKST (the
6QZ sample).  The 2dF instrument is a multi-fibre spectrograph which
can obtain simultaneous spectra of up to 400 objects at once over a
$2^{\circ}$ diameter field of view, and is positioned at the prime
focus of the Anglo-Australian Telescope.  Fibres are robotically
positioned within the field of view and are fed to two identical
spectrographs (200 fibres each).  Two field plates, and a tumbling
system allow one field to be observed while a second is being
configured, reducing down-time between fields to a minimum.  The
spectrographs each contain a Tektronix $1024\times1024$ CCD with
24~$\mu$m pixels.  Details of the 2dF instrument can be found in Lewis
et al. (2002).

The 6dF instrument is a multi-fibre spectroscopic instrument on the
UKST with the capability to simultaneously observe up to 150 sources
over a $6^{\circ}$ diameter field of view.  Fibres are positioned
robotically onto a field plate, which is then installed at the focus
of the UKST telescope.  Fibres are fed to a bench mounted spectrograph
containing an EEV $1032\times1027$ CCD with 13~$\mu$m pixels.  Further
details can be found in Watson et al. (2000).

Based on the need to restrict the magnitude range of objects
observed by 2dF (to reduce scattered light) and the observed surface
density of candidates, objects in the input catalogue with $18.25 <
\bj \leq 20.85$ were observed with 2dF and those with  $16.0 < \bj \leq
18.25$ were observed with 6dF.  Although the 6QZ sample is
approximately 100 times smaller than the 2QZ, it is important in
extending the  coverage of the QSO OLF to almost a factor of 100 in
luminosity based on a single homogeneous input catalogue.

Full details of the 2dF spectroscopic observations and subsequent
catalogue were provided by Croom et al.\ (2001b) at the time of the
initial 10k data release.  We therefore only present a brief
description in this paper, updating information where appropriate.
2dF spectroscopic observations began in January 1997 and were
completed in April 2002.  The spectral dispersion was
4.3\AA~pixel$^{-1}$, giving an instrumental resolution of 9\AA.  
The spectra covered the wavelength range
3700--7900\AA.  Each 2dF field in the survey was observed for 3300 --
3600 secs, giving a median signal-to-noise ratio ($S/N$) of 5.0 per
pixel in the $B$-band (4000--4900\AA), see
Fig. \ref{fig:sn}a. The data were taken in a wide variety of
conditions. Under conditions of poor seeing ($> 2\,$arcsec) or
transparency, exposure times were increased to compensate for the
lower signal rates.   In the event that conditions prevented any field
reaching its pre-determined completeness level, it was scheduled for
re-observation (see below).

The 2QZ input catalogue was merged with that of the 2dF Galaxy
Redshift Survey (2dFGRS; Colless et al. 2001) and a complex tiling
algorithm was applied to the resultant joint catalogue in order to
maximize the efficiency with which the two declinations strips could
be covered with the minimum number of circular 2dF fields.  The 2QZ
survey area is an exact subset of the 2dFGRS area.  Where possible,
when conditions were unsuitable for the more exacting requirements of
the 2QZ program (e.g. poor seeing, significant moonlight),
observations of the 2dFGRS only fields were carried out.

6dF observations were performed over the period March 2001 --
September 2002. The spectral range covered was 3900--7600\AA\ using a
low dispersion 250B grating which provided a dispersion of
286\AA~mm$^{-1}$ (3.6\AA~pixel$^{-1}$) and a spectral resolution of
$\simeq11.3$\AA.  Typical observation times were between 1.5 hours and
2 hours per field.  An average of 50 6QZ objects were observed in each
6dF pointing.  A median $S/N=16$ per pixel was obtained in the continuum of 6dF
spectra (see Fig. \ref{fig:sn}b).  This higher $S/N$ was set by our
requirement to effectively identify the much larger numbers of
contaminating stars in the 6QZ.  A number of 6QZ targets where also
observed in preliminary survey work carried out between September 1996
and December 1999 with the FLAIR spectrograph \cite{flair94} on the
UKST prior to the commissioning of 6dF.  These FLAIR spectra cover the
range 3800--7300\AA\ using the same 250B grating with $\simeq11$\AA\
resolution, but are generally lower $S/N$ than the 6dF data.  All but
a few (28) QSOs found with FLAIR have been re-observed with 6dF.
FLAIR spectra are not available as part of the publically available
database.  Observations were obtained for candidates in both
declination strips.  However, the 6dF observations of the equatorial
strip candidates are quite incomplete (less than 60 per cent of the
candidates were observed), and so only the SGP strip
is included in the final catalogue.  In total, 25 overlapping 6dF
pointings were used to cover the SGP strip.

\subsection{Catalogue composition}

\begin{table*}
\caption{2QZ/6QZ catalogue composition as a function of minimum sector
completeness level (see text).  The 70, 80 and 90 per cent columns
give the properties of survey sub-samples which have those
spectroscopic completenesses respectively.  We list the number of
objects in each of the different quality classes (Q1, Q2 and Q3).
The final row gives the fraction (compared to the total number of {\it
observed} sources) of IDs in each of the specified sub-samples and
quality classes.} 
\label{tab:numbers}
\begin{tabular}{@{}lrrrrrrrrrrrrrrr@{}}
\hline
&\multicolumn{12}{c}{2QZ}&\multicolumn{3}{c}{6QZ}\\
&\multicolumn{3}{c}{All}&\multicolumn{3}{c}{70 per cent}&\multicolumn{3}{c}
{80 per cent}&\multicolumn{3}{c}{90 per cent}&\multicolumn{3}{c}{All}\\
Class.&  Q1  &  Q2  &  Q3  &  Q1   &  Q2  &  Q3  &  Q1   &  Q2  &  Q3  &  Q1   &  Q2  &  Q3  &  Q1  &  Q2  &  Q3  \\
\hline 
QSO  & 22655 & 683  &  --  & 20905 &  480 &  --  & 18068 &  312 &  --  & 11046 &  86  &  --  &  317 &  5   &  --  \\
NELG & 4484  & 74   &  --  & 4086  &   50 &  --  & 3458  &  31  &  --  & 2005  &  10  &  --  &   50 &  0   &  --  \\
gal  & 79    & 16   &  --  & 73    &   11 &  --  & 59    &   6  &  --  & 34    &   1  &  --  &    1 &  0   &  --  \\
star & 10904 & 1388 &  --  & 9965  & 1007 &  --  & 8466  & 676  &  --  & 4996  & 168  &  --  & 1148 &  4   &  --  \\
cont & 102   & 52   &  --  & 84    &   41 &  --  & 68    &  28  &  --  & 24    &  12  &  --  & 3    &  1   &  --  \\
??   &  --   &  --  & 4139 &  --   &  --  & 2749 &   --  &  --  & 1669 &  --   &  --  & 405  &  --  &  --  &  35  \\
Total& 38224 & 2213 & 4139 & 35113 & 1589 & 2749 & 30119 & 1053 & 1669 & 18105 &  277 & 405  & 1519 &  10  &  35  \\
ID Frac&0.858& 0.050& 0.093& 0.890 & 0.040& 0.070& 0.917 & 0.032& 0.051& 0.964 & 0.015& 0.022&0.971 &0.006 & 0.022\\
\hline
\end{tabular}
\end{table*} 
\subsubsection{The 2QZ catalogue}\label{sec:2qzcat}

2dF data were reduced at the telescope using the 2dF data-reduction
pipeline {\small 2DFDR} \cite{2dfdr,2dfman} and information on the
spectroscopic completeness achieved on each field was immediately fed
back into the catalogue.  2dF fields which failed to achieve a
pre-determined completeness level (either 70 per cent for the 2QZ or
90 per cent for the 2dFGRS) were scheduled for re-observation.  In the
end 556 out of 573 fields in the two 2QZ declination strips were
observed.  These included 10 fields that were kindly added to the
equatorial strip when observations of 2QZ (and 2dFGRS) sources were
made using spare fibres from spectroscopic follow up of the Millennium
Galaxy Catalogue \cite{mgc}.

The coverage map for the 2dF survey is given in Fig. \ref{fig:cover},
showing the percentage of 2QZ targets that we obtained spectra for.
We failed to obtain observations of a few 2dF fields in the most
Northerly declination band of the equatorial strip at RAs greater than
14$^{\rm h}$ and a few fields at the very extreme edges/corners of
both strips.  In total, spectra were obtained for 44576 objects (93.3
per cent of the input catalogue) in the 2QZ, corresponding to an
effective area of 673.4$\,$deg$^{2}$.  A small
number of objects in the catalogue were observed not by 2dF, but by
other instruments and/or telescopes.  These are also included in our
final catalogue.  111 equatorial sources with NVSS \cite{nvss} radio
detections and $\bj<20$ were observed with the Low-Resolution Imaging
Spectrograph on the Keck II Telescope \cite{brotherton98}.  Also, 69
targets were observed with either the Double-Beam Spectrograph (DBS)
on the Siding Spring Observatory 2.3m Telescope, or the RGO
Spectrograph on the AAT, as part of the follow up of close ($<20''$)
QSO candidate pairs \cite{smith98}.  DBS spectra of 20 objects in
the 6QZ were also obtained.

Objects without spectra in the 2QZ either lie in the
fields that were not observed by 2dF or in regions of high 2QZ/2dFGRS
candidate surface density where the tiling algorithm was unable to
configure a fibre observation efficiently given the minimum
fibre-to-fibre spacing restriction ($\sim 30\,$arcsec).  However, we
note that 2QZ targets were given higher priority in the tiling
algorithm than 2dFGRS targets, in order not to imprint the stronger
angular clustering pattern of the galaxies onto the QSO distribution.
Due to the significant amount of overlap between 2dF fields and the
repeat observations of low completeness fields, 10528 objects (23.6
per cent) in the 2QZ have more than one spectroscopic observation.  

Once reduced, 2QZ spectra were classified using the {\small AUTOZ}
program (see Croom et al.\ 2001b) which uses a $\chi^2$-minimization
technique to fit each spectrum to a number of QSO (BAL and non-BAL),
galaxy and stellar templates and measure a redshift for all
extra-galactic identifications.  The QSO template was based on the
composite spectrum of Francis et al. (1991).  {\small AUTOZ} produces
a single identification based on the best fit template in one of six
categories based on the following spectral criteria:
\begin{tabbing}
1\=12345678\= \kill
\>{\bf QSO: } \> Broad ($> 1000\,$\kms) emission lines.\\
\>{\bf NELG:} \> Narrow ($< 1000\,$\kms) emission lines only.\\
\>{\bf gal: } \> Redshifted galaxy absorption features.\\
\>{\bf star:} \> Stellar absorption features at rest.\\
\>{\bf cont:} \> No emission or absorption features (High S/N).\\
\>{\bf ??:}   \> No emission or absorption features (Low S/N).
\end{tabbing}

All {\small AUTOZ} identifications were independently checked
visually by at least two members of the 2QZ team to correct
any identifications that were clearly in error.  Approximately 5 per
cent of the {\small AUTOZ} identifications were changed in this
fashion.  As part of the classification process a quality flag
was attached to each identification and redshift measurement as
follows: 
\begin{tabbing}
1\=1234567890123\= \kill
\>{\bf Quality 1:} \>  High-quality identification or redshift.\\
\>{\bf Quality 2:} \>  Poor-quality identification or redshift.\\
\>{\bf Quality 3:} \>  No identification or redshift assignment.\\
\end{tabbing}

The quality flag was determined independently for the identification
and redshift of an object.  For example, a quality 1 QSO
identification could have a quality 1 or 2 redshift. 

\begin{figure}
\vspace{-0.5cm}
\centering 
\centerline{\psfig{file=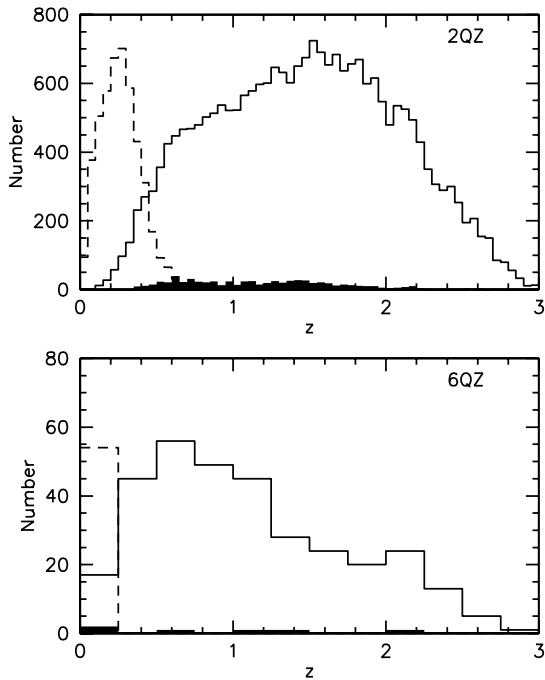,width=10cm}}
\caption{Number-redshift histograms for the 2QZ (upper panel) and 6QZ
(lower panel) showing both QSOs (solid lines) and NELGS (dashed
lines).  The unshaded solid line includes all quality 1 and 2 QSOs,
while the shaded histograms denote quality 2 QSO identifications
only. Bin widths are $dz=0.025$ and $dz=0.25$ for the 2QZ and 6QZ 
respectively.} 
\label{fig:nzcat}
\end{figure}

\begin{figure}
\vspace{-0.5cm}
\centering 
\centerline{\psfig{file=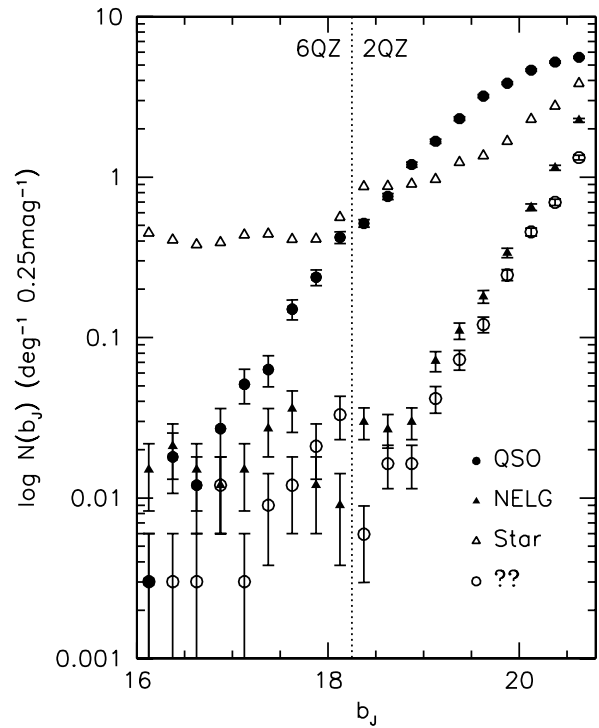,width=10cm}}
\caption{Number-magnitude counts for QSO (filled circle), NELG (filled
  triangle), Star (open triangle) and ?? (open circle) classifications
  in the 2QZ and 6QZ. Poisson error bars are shown.}
\label{fig:nbcat}
\end{figure}

Table \ref{tab:numbers} gives the breakdown between the numbers
identified in each class in the 2QZ. Fig. \ref{fig:nzcat} shows the
number-redshift relation, $n(z)$, for the QSOs and NELGs in the full
2QZ.  The $n(z)$ histograms for both classes of objects  are
relatively smooth, indicating that there are no strong biases towards
the selection of QSOs/NELGs over specific redshift intervals.  The
decline in QSO numbers at $z<0.4$ and $z>2.2$ are largely due to
incompleteness in the colour and morphological selection process (see
Section  \ref{sec:comp}).  We show the number counts, $n(\bj)$, for
each identification class (omitting the small numbers of 'cont' and
'gal' identifications for clarity) in Fig \ref{fig:nbcat}.  The
surface densities are based on the observed counts divided by
effective area of the full survey, with no further correction for
incompleteness.  At magnitudes fainter than $\bj\sim18.5$ mag the QSOs
are the largest population, while at brighter magnitudes, galactic
stars dominate.

The 'cont' identification relates to potential BL Lac sources, however
this class of object could be contaminated by other weak-lined objects
(e.g. DC white dwarfs).  To indicate the uncertainty inherent in this
classification, in our final catalogue all 'cont' IDs have quality 2.
A more detailed investigation of possible BL Lacs in the survey has
been carried out by Londish et al. (2002).

\begin{figure*}
\vspace{-0.5cm}
\centering 
\centerline{\psfig{file=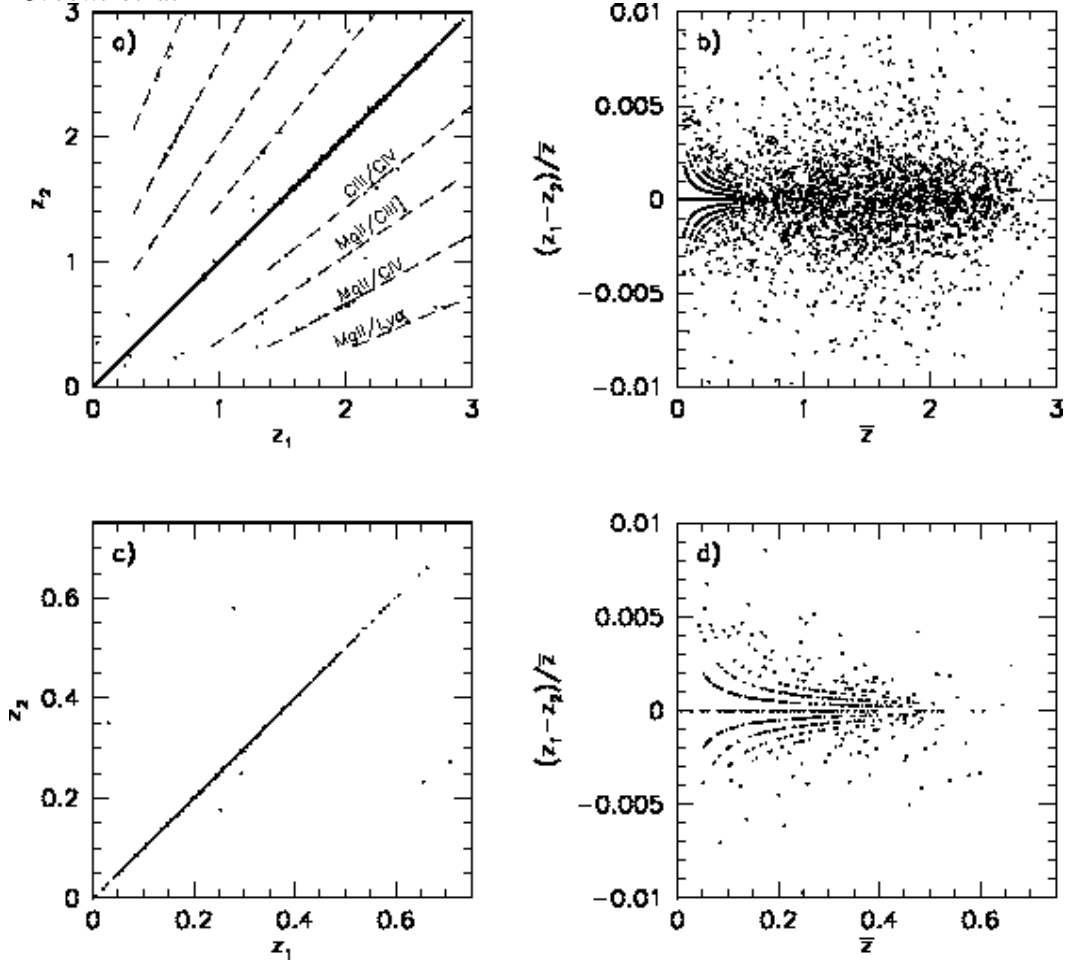,width=14cm}}
\caption{Redshift comparison for extra-galactic emission line objects
in the 2QZ.  a) $z_1$ vs. $z_2$ for QSOs.  The dashed lines indicate
the redshift differences caused by the confusion of specific QSO
emission lines.  b) Relative difference vs. mean $z$ for QSOs.  
c) $z_1$ vs. $z_2$ for NELGs.  d) Relative difference vs. mean $z$ for
NELGs.  The quantization seen in b) and d) at low redshift is due to
the redshifts being determined to a precision of 4 decimal places
($\sim 30$\kms).  In all cases the determined redshift errors are
considerably larger than this.}
\label{fig:zcomp}
\end{figure*}

Of the 44576 objects with spectroscopic observations in the full 2QZ,
38224 received quality 1 IDs (85.8 per cent) 2213 quality 2 IDs (5.0
per cent) and 4139 quality 3 IDs (9.3 per cent).
 
We obtained a quantitative assessment of the quality flags from the
objects with repeat observations. Based on the 5073 objects with
quality 1 identifications and redshifts in two observations; 186 (3.7
per cent) had a different identification and different redshift. A
further 38 (0.75 per cent) objects had a different identification, but
the same redshift (to within $\Delta z=0.015$); arising as a result of
a classification change from NELG to QSO or vice versa.  2697 objects
with quality 1 IDs were classified as QSOs in both the observations,
of these 2589 (96.0 per cent) had the same redshifts.  Amongst the
615 objects classified as NELGs in both observations, 609 (99.0 per
cent) were assigned the same redshift.

Of the 1026 objects with a quality 1 identification and redshift, but
also a quality 2 identification, 334 (32.6 per cent) had a different
identification and redshift.  21 (2.0 per cent) had a different
identification but the same redshift.  Of the 265 (14) objects that
were classified as QSOs (NELGs) in both observations  185 (12) had the
same redshift, or 70 (86) per cent of the respective populations. 

Overall, we therefore conclude that the quality 1 identifications and
redshifts are more than 95 per cent reliable, while the quality 2
identifications and redshifts are only reliable at the 70 per cent
level.

We also obtained an estimate of the likely composition of the quality
3 object catalogue by looking at the 2735 quality 3 objects that were
subsequently re-observed and given a quality 1 identification and
redshift.   Of these objects, 1611 (58.9 per cent) were classified as
QSOs, 902 (33.0 per cent) as stars, 212 (7.8 per cent) as NELGs and 10
(0.4 per cent) as gals.  This is close to the distribution of
classifications in the full 2QZ (excluding ?? objects, see Table
\ref{tab:numbers}), with a slightly higher fraction (4 per cent) of
stars and a lower fraction of NELGs amongst the unidentified objects.
QSOs remain at an approximately constant 59 per cent of sources
amongst both the identified and unidentified populations. 

We can also increase the mean completeness of the 2QZ catalogue by
limiting the sectors formed from the overlap of  individual 2dF
observations included in the catalogue to  only those which meet a
specified spectroscopic completeness  threshold (see Section
\ref{sec:comp} below).  Table \ref{tab:numbers} lists the composition
of the catalogue based on a number of spectroscopic completeness
thresholds set at 70 per cent, 80 per cent and 90 per cent (based on
the percentage of quality 1 identifications in a given sector).  By
spectroscopic completeness we mean the fraction of spectroscopically
observed sources for which we obtained a quality 1 identification.  In
most  analysis that we have carried out with the 2QZ, we  have used
the 70 per cent sector spectroscopic completeness threshold to define
the sample.  Furthermore, we have only used objects with a quality 1
identification.  These criteria were chosen to give the best
compromise between maximizing the sample size (over 20000 QSOs) and
minimizing spectroscopic incompleteness (11 per cent) in the 2QZ.

The repeat observations also provide a useful method to determine
redshift accuracy.  The RMS pairwise dispersion between redshift
measurements for the 2589 QSOs with quality 1 redshifts and $\Delta z
< 0.015$ (excluding the objects with an incorrect redshift due to line
mis-identification) is $\sigma(\Delta z)=0.0038$; giving a mean
redshift error of $\sigma(z)=0.0027$ for an individual measurement.
We note, however, that the dispersion increases as a function of
redshift, from $\sigma(z,z<1)=0.0022$ to $\sigma(z,z>2)=0.0044$.  A
more z-independent estimate of the redshift error  is given by the
fractional error in the redshift $\sigma(z)/z=0.0027$, which is
actually identical to the mean of $\sigma(z)$.  A plot of the
redshifts for the repeat observations with quality 1 identifications
and redshifts is shown in Fig. \ref{fig:zcomp}.  From this we can also
see that the most common mis-identification was between the \civ\
$\lambda1549$ and \mgii\ $\lambda2798$ lines.  For NELGs, the RMS
dispersion in the redshifts $\sigma(\Delta z)=0.0005$, corresponding
to a mean error in an individual redshift measurement error of
$\sigma(z)=0.00035$.

\subsubsection{The 6QZ catalogue}

\begin{table*}
\centering
\caption{Format for the 2QZ catalogue.  The format entries are based
on the standard {\small FORTRAN} format descriptors.}
\label{tab:cat}
\begin{tabular}{@{}lrl@{}}
\hline
Field&Format&Description\\
\hline
Name     &          a20 & IAU format object name\\
RA       &   i2 i2 f5.2 & RA J2000 (hh mm ss.ss)\\
Dec      & a1i2 i2 f4.1 & Dec J2000 ($\pm$dd mm ss.s)\\
Catalogue number &   i5 & Internal catalogue object number\\
Catalogue name   &  a10 & Internal catalogue object name\\
Sector   &          a25 & Name of the sector this object inhabits\\
RA       &   i2 i2 f5.2 & RA B1950 (hh mm ss.ss)\\
Dec      & a1i2 i2 f4.1 & Dec B1950 ($\pm$dd mm ss.s)\\
UKST field &         i3 & UKST survey field number\\
X{\rm APM} &       f9.2 & APM scan X position ($\sim8$ $\mu$m pixels)\\
Y{\rm APM} &       f9.2 & APM scan Y position ($\sim8$ $\mu$m pixels)\\
RA       &        f11.8 & RA B1950 (radians) \\
Dec      &        f11.8 & Dec B1950 (radians) \\
$\bj$    &         f6.3 & $\bj$ magnitude\\
$u-\bj$  &         f7.3 & $u-\bj$ colour\\
$\bj-r$  &         f7.3 & $\bj-r$ colour [including $r$ upper limits
as: $(\bj-r_{\rm lim})-10.0$]\\
$N_{obs}$&           i1 & Number of observations\\
\hline
Observation \# 1\\
$z_1$    &         f6.4 & Redshift\\
q$_{1}$  &           i2 & Identification quality $\times$ 10 + redshift quality\\
ID$_1$   &          a10 & Identification\\
date$_1$ &           a8 & Observation date\\
fld$_1$  &           i4 & 2dF field number $\times 10$ + spectrograph number\\
fibre$_1$&           i3 & 2dF fibre number (in spectrograph)\\
$S/N_1$  &         f6.2 & Signal-to-noise ratio in 4000--5000{\AA} band\\
\hline
Observation \# 2\\
$z_2$    &         f6.4 & Redshift\\
q$_{2}$  &           i2 & Identification quality $\times$ 10 + redshift quality\\
ID$_2$   &          a10 & Identification\\
date$_2$ &           a8 & Observation date\\
fld$_2$  &           i4 & 2dF field number $\times 10$ + spectrograph number\\
fibre$_2$&           i3 & 2dF fibre number (in spectrograph)\\
$S/N_2$  &         f6.2 & Signal-to-noise ratio in 4000--4900{\AA} band\\
\hline
$z_{\rm prev}$&    f5.3 & Previously known redshift (V\'eron-Cetty \& V\'eron 2000)\\
radio    &         f6.1 & 1.4GHz Radio flux, mJy (NVSS) \\
x-ray    &         f7.4 & X-ray flux, $\times10^{-13}\,$erg$\,$s$^{-1}$cm$^{-2}$ (RASS)\\
dust     &         f7.5 & $E(B-V)$ (Schlegel et al. 1998)\\
comment$_1$ &       a20 & Specific comments on observation 1\\
comment$_2$ &       a20 & Specific comments on observation 2\\
\hline
\end{tabular}
\end{table*}

An identical procedure was followed to reduce the 6dF data and carry
out the spectroscopic identification of the 6QZ.  In total, 6dF
spectra were obtained for 1564 of the 1657 colour-selected candidates
(94.4 per cent) in the SGP strip, giving an effective
area of 333.0$\,$deg$^2$.

The composition of the 6QZ catalogue is given in Table
\ref{tab:numbers}.  The 6QZ contains a much higher fraction of
galactic stars (73.7 per cent of observed sources) and a
correspondingly lower fraction of  QSOs (20.6 per cent) given its
brighter magnitude limit.  The number-redshift relation for the  QSOs
and NELGs in the 6QZ is shown in Fig. \ref{fig:nzcat}.   The $n(\bj)$
relation is shown in Fig. \ref{fig:nbcat}.  We note that the QSO
$n(\bj)$ relation fits smoothly onto that obtained for the 2QZ at
$\bj=18.25$, while the stellar $n(\bj)$ relation obtained for the 6QZ
lies approximately a factor 2 lower than expected from an
extrapolation of the 2QZ stellar counts.  This reflects the more
restrictive colour cut, $(u-\bj)\leq-0.50$, used to select the 6QZ
candidate list.  At this cut, the vast majority of the stars
identified have been scattered into this sample by photometric errors.
This explains the increase in the stellar surface density at the very
brightest magnitudes ($\bj < 17\,$mag), where the photometric errors
on the photographic magnitudes increase.

The 6QZ also has a much higher spectroscopic completeness level (over
97 per cent for quality 1 identifications) than the 2QZ and so we
choose to make no further cuts on the basis of sector completeness as
we did for the 2QZ.  The higher completeness levels are directly
attributable to the much higher mean S/N obtained for the 6dF
observations.

Repeat observations were obtained for 475 objects in the 6QZ.  Based
on a similar comparison of the quality 1 identifications and redshifts
to that carried out for the 2QZ sample, we conclude that the quality 1
identifications in the 6QZ are 99 per cent accurate (only 2 out of 254
repeat quality 1 identifications changed between observations) and the
quality 1 redshifts are over 96 per cent accurate (only 6 out of 177
QSO  redshift measurements differed by more than $\Delta z =0.015$,
and of these, only one QSO had $\Delta z > 0.03$).

We obtain a redshift error of $\sigma(z) = 0.0026$ for the 6QZ QSOs,
identical to that obtained for the 2QZ, based on a comparison of the
redshifts obtained for the 171 QSOs in the 6QZ with repeated  quality
1 observations with $\Delta z<0.015$.

\subsection{Data Products}

The 2QZ and 6QZ catalogues are available as ASCII files from the  2QZ
WWW site, {\tt http://www.2dfquasar.org}, and on a CD-ROM release.
The catalogue format is identical for both surveys and is given in
Table~\ref{tab:cat}.  Note that the catalogue format has been extended
to include more information since the preliminary data release
\cite{2qzpaper5}.  The first part of a catalogue entry contains
details from the input catalogue, such as position and magnitude.  We
note that the object names may in some cases not correspond exactly to
the source position in $\alpha$ and $\delta$ as we have improved the
astrometry of the catalogue \cite{2qzpaper3} since the object names
were defined.  The names have not been changed subsequently, to avoid
confusion with previously published lists.  We now include internal
catalogue names and numbers, although objects should generally be
referenced by their main IAU format name.  A new entry is included for
the name of the sector inhabited by the object.  A sector is defined
as the intersection of overlapping 2dF fields (see Section
\ref{sec:comp}).  The format of these is e.g. S\_200\_201\_247, where the
S denotes the SGP strip and the numbers indicate that the sector is
formed by the overlap of fields 200, 201 and 247.  There are no
sectors defined for the 6QZ.  Coordinates are also supplied in the
B1950 system, as the survey was constructed in this system and the
production of completeness masks etc. is more straight forward in this
system.  We also note that sources which had only upper limits
(i.e. non-detections) on the $r$ plates are also included in the
catalogue and have a listed $\bj-r$ colour.  In this case the colour
term is $(\bj-r_{\rm lim})-10$, the $-10$ being used to differentiate
upper limits from normal colours (objects with real $r$-band
detections have colours in the range $-1.4<\bj-r<3.4$, while upper
limits have $\bj-r<-9.8$).

The input catalogue information is followed by details of observations
and identifications.  As discussed above, in a number of cases we have
two or more 2dF observations of the same source.  The first and second
observations (arranged in quality order) are listed in the main
catalogue.  These are useful to assess the quality of  the final
catalogue.  The identification and redshift that we adopt is that from
the observation with the numerically lowest quality value (where the
quality value is identification quality$\times10$ + redshift quality),
and if there are equal quality values, the highest S/N.  In all cases
the final adopted ID is listed as observation \#1, with the lower
quality observation being listed as observation \#2.  We also list the
2dF field number (fld$_1$ and fld$_2$) for each observation.  These
correspond to the actual 2dF field number $\times 10$ + the
spectrograph number (1 or 2 for 2dF, 6 for 6dF).  Sources with spectra
from the Keck observations of 2QZ radio sources all have field number
8881.  Sources from the followup of close pairs have field number
7771.  Sources from the 6QZ with DBS observations have field number
6666.  We also include cross-matches to selected other databases.
Where a 2QZ/6QZ source matches the position (to within $6''$) of a
previously known QSO/AGN in the catalogue of V\'eron-Cetty \& V\'eron
(2000) we include the previously known redshift.  We also include
radio fluxes at 1.4GHz from the NRAO VLA Sky Survey (NVSS; Condon et
al. 1998) and X-ray fluxes from the ROSAT All Sky Survey (RASS; Voges
et al. 1999; Voges et al. 2000), converting from RASS counts per
second to flux (in erg$\,$s$^{-1}$cm$^{-2}$) by multiplying by
$1.2\times10^{-11}$.  The matching radii used were $15''$ and $30''$
for NVSS and RASS respectively.  An estimate of the galactic reddening
$E(B-V)$ to  each source is taken from the work of Schlegel,
Finkbeiner \& Davis (1998).  In order to convert from $E(B-V)$ to
extinction values multiply by 5.434, 4.035 and 2.673 for $A_{\rm u}$,
$A_{\rm b_{\rm J}}$ and $A_{\rm r}$ respectively.  Lastly we supply
any comments made on the particular observation which might flag the
object or spectrum as unusual.


For the catalogue we have not attempted to sub-divide the NELG class
into LINERs, Seyferts, starbursts, etc. Given the finite spectral
window of the 2QZ and 6QZ observations and the large redshift range
observed for NELGs ($0<z<0.6$), in many case key diagnostic lines
required for further spectral classification are missing from the
spectra.  The spectral coverage provided by 2dF and 6dF limits our
ability to detect broad \mgii\ emission until $z>0.35$.  In some QSOs
with $0.35<z<0.5$, broad \mgii\ is clearly detected with no
corresponding broad H$\beta$ seen.  Thus at lower redshifts,
$0.15<z<0.35$, some objects classified as NELGs on the basis of a
narrow H$\beta$ line may exhibit broad \mgii\ below the blue limit of
the 2dF spectral window.  For QSOs with $z>1.6$, {\small AUTOZ} fits
both `normal' and BAL QSO templates.  Objects identified as BALs are
indicated by `QSO(BAL)' in the identification column.  However, due to
the varied nature of BAL QSOs, this will not be a comprehensive list
of BALs.  A more detailed search for BALs has not yet been applied.

Within the objects we classify as stars, in the main we only give
classifications for white dwarfs (WDs).  We have not attempted to
provide spectral classification of main sequence stars.  The
$\sim2000$ WDs are mostly DAs with strong broad hydrogen Balmer
absorption lines.  At low S/N these can be confused with A stars.
Only detailed fitting of the temperature and gravity of these stars
can easily resolve this issue, and this is beyond the scope of the
present paper.  As well as DAs we also find a number of DB and DO WDs
which are dominated by neutral and singly ionized helium respectively.
A small number of DZs are also found, with broad \caii~H and K
$\lambda\lambda3969,3933$  absorption.  The only other stellar types
we classify are emission line stars with strong hydrogen Balmer
emission which we classify as CV (cataclysmic variable) and WD + M
dwarf binary systems which are denoted as DAM, DBM etc.  We note that
our code has been designed primarily to simply identify an object as a
star (and therefore not a QSO), and that a more detailed analysis
would no doubt provide more accurate classifications.  Some analysis
of the Galactic stars in the 2QZ has already taken place
\cite{vennes02,marsh02}.  The above classifications are denoted within
the catalogue in parentheses after the main star identification
[e.g. star(DB), star(CV)].

FITS format spectra for all objects, including repeat observations and
spectra from which no identification was possible, are provided as a
primary data product of the 2QZ and 6QZ surveys at {\tt
http://www.2dfquasar.org/} and on the survey CD-ROM.  Note that many
of the repeat spectra are of low quality, as the fields were repeated
if they had low completeness.  A small fraction of spectra have bad
'fringing' caused by a damaged fibre. This shows up as a strong
oscillation as a function of wavelength, rendering the underlying
spectrum unusable.  In a few cases {\small 2DFDR} failed to cope
adequately with the sky subtraction process at either end of the CCD
data-frame and so a number of spectra have poorly subtracted night sky
emission (and in some cases moonlight) which has also affected our
ability to identify an object in some cases.  The spectral database
does not include spectra from FLAIR observations.  It also does not
include spectra from the small number of 2QZ identifications not based
on 2dF spectra (see Section \ref{sec:2qzcat}).

On the WWW site, we also provide 1-arcmin square SuperCOSMOS-scanned
images \cite{supercosmos} of the $J$ photographic plates
around each object in the 2QZ and 6QZ catalogue. Due to space
restrictions, these images are not available on the CD-ROM.  The input
catalogue is also available from this site \cite{2qzpaper3}. The input
catalogue, spectroscopic catalogue, FITS spectra and photographic
images define the primary data products of the 2QZ and 6QZ surveys.
To aid in the interrogation and analysis of the catalogue, we also
provide a web search tool, scripts/code used to generate the
survey mask for an arbitrary minimum sector completeness and the
$\alpha,\delta$ of the survey boundaries/holes drilled around bright
stars in the input catalogue.  We also provide quantitative estimates
of catalogue completeness for QSOs, where appropriate, as a function
of magnitude, redshift and position in 2QZ and 6QZ catalogues.  The
empirical methods used to define these completeness estimates, arising
from a variety of sources are discussed in turn in the following
section.

\section{Survey completeness}\label{sec:comp}

\begin{figure*}
\vspace{-0.5cm}
\centering 
\centerline{\psfig{file=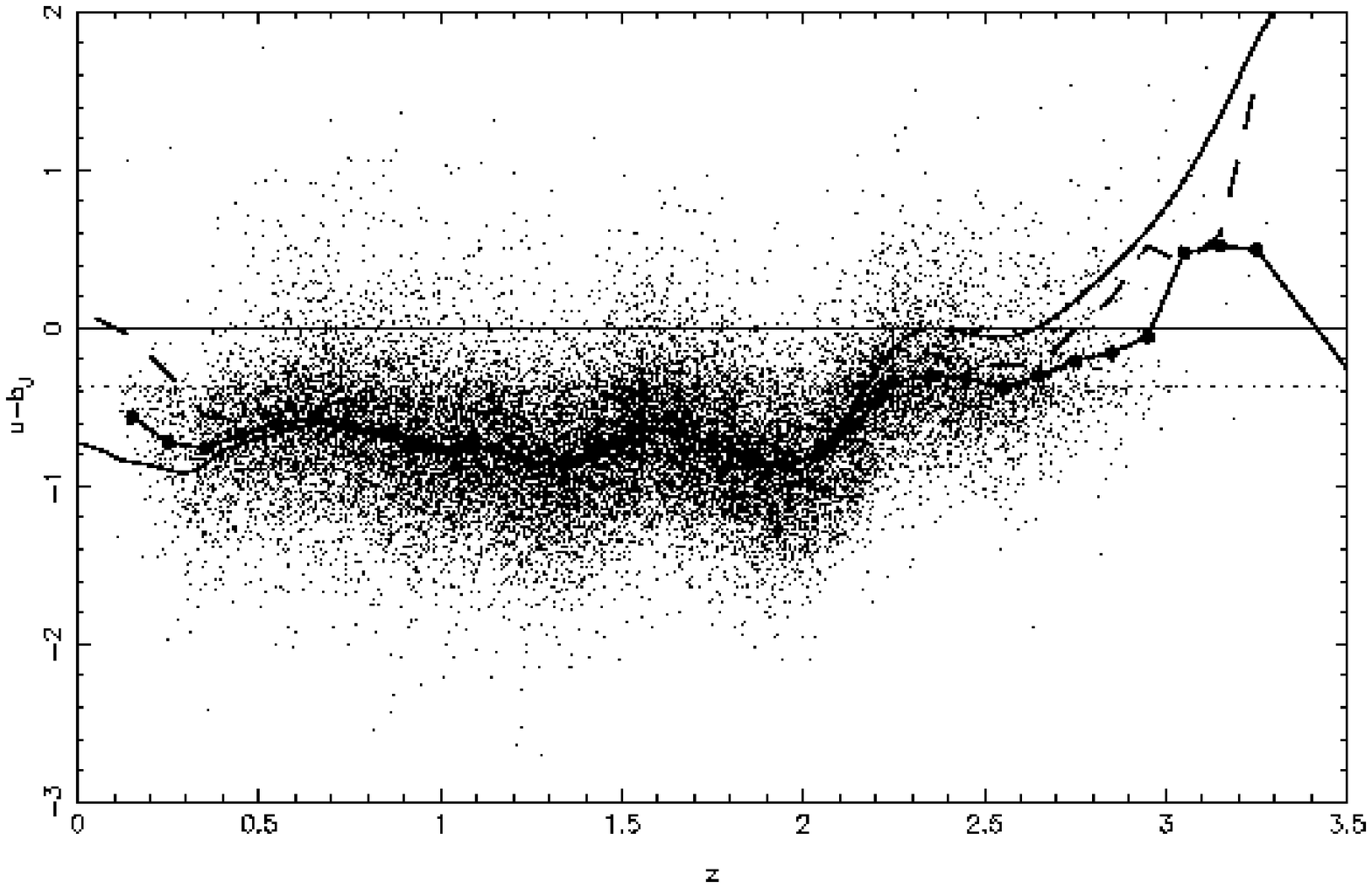,width=16.0cm}}
\centerline{\psfig{file=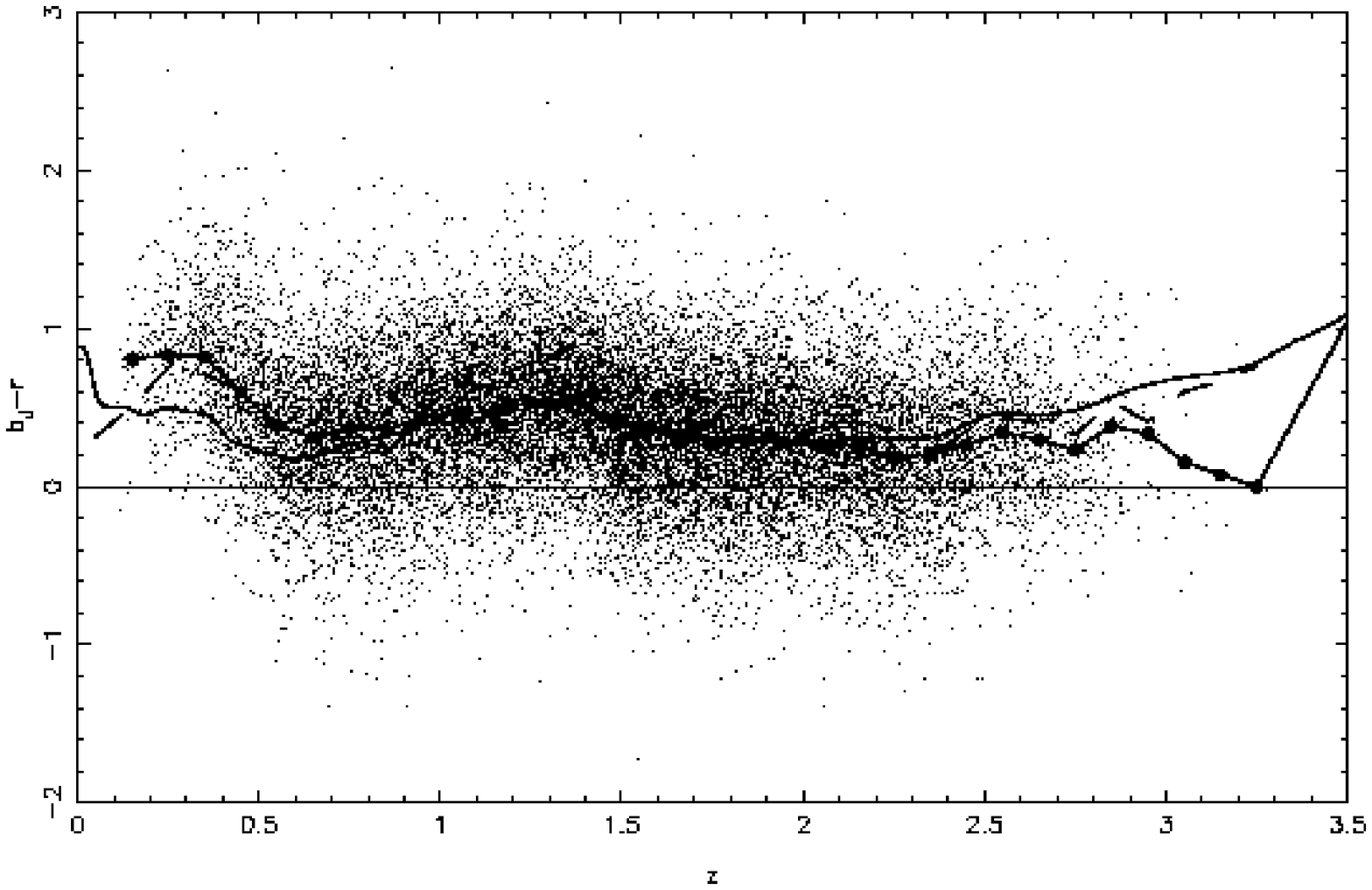,width=16.0cm}}
\caption{The colour redshift relations for the 2QZ/6QZ sample.  Top:
$u-\bj$ vs. $z$ for all quality 1 QSOs.  The dotted line indicates the
$u-\bj=-0.36$ selection limit.  Bottom:  $\bj-r$ vs. $z$ for quality 1
QSOs.  The connected filled circles show the mean colours in $\Delta
z=0.1$ bins.  The solid line describes the colours derived from the
SDSS composite spectrum, zero-pointed to the 2QZ at $z=1-2$.  The
dashed line describes the mean colours of the Hawkins (priv. comm.)
variability selected sample.}
\label{fig:colz}
\end{figure*}

We now come to discuss the survey completeness, and sources of
incompleteness.  We define four separate types of {\it completeness}
that can be used to describe the 2QZ and 6QZ surveys.  These are:
\begin{itemize}
\item {\bf Morphological completeness}, $f_{\rm m}(\bj, z)$.
  This describes our effectiveness at differentiating between point sources
  and extended sources on UKST plates, and also the possible effect of
  the QSO host galaxies on morphology at low redshift. 
\item {\bf Photometric completeness}, $f_{\rm p}(\bj, z)$.  This
  attempts to take into account any QSOs which may have fallen outside
  of our colour selection limits.  
\item {\bf Coverage completeness (or coverage)}, $f_{\rm c}(\theta)$.
  We define the coverage as the fraction of 2QZ catalogue
  sources which have spectroscopic observations.
\item {\bf Spectroscopic completeness}, $\fs(\bj, z, \theta)$.
  This is the fraction of 2QZ catalogue objects which have
  the specified quality of spectrum (in our case we consider quality 1
  identifications).
\end{itemize}
We indicate against each type of completeness above whether it is
considered to be a function of angular position $\theta$, magnitude
$\bj$, or redshift $z$.  To provide a general description of the
angular positional dependence of completeness we determine
completeness values within each of the regions which are described by
sets of our overlapping $2^{\circ}$ diameter fields.  Within each of
these regions, which we call sectors, we may define a coverage and
spectroscopic completeness value.  For ease of use these completeness
maps are then pixelized on a scale of 1 arcminute.  Below we describe
each of these different  completeness measurements and the corrections
which can be applied to account for them.  For some analyses it is
appropriate to define such completeness on an object-by-object basis.

\subsection{Morphological completeness}

Morphological completeness falls into two categories.  The first
concerns objects which are truly point-sources, but which have been
mis-classified by the APM analysis software.  The second concerns QSOs
which have significant host galaxy contributions, which make their
images non-stellar.  This is an issue at low redshift ($z<0.4$).

The first of these two classes of morphological completeness is by far
the easiest to quantify.  A small fraction of stellar objects are
classified as merged images or galaxies by the APM on the basis of
random errors or overlapping images on the photographic plate.  The
fraction of sources  classified as non-stellar ($f_{\rm ns}$) and lost
in this fashion has been determined by direct comparison with the SDSS
Early Data Release \cite{sdssedr} imaging data in the equatorial
region where the surveys overlap \cite{2qzpaper3} and is well
described by
\begin{equation}
f_{\rm ns}=0.064+0.0052(\bj-16).
\end{equation}
$f_{\rm ns}$ thus varies between 0.064 and 0.089 over the full
magnitude range covered by the 2QZ and 6QZ surveys.  We see no
systematic variations in $f_{\rm ns}$ as a function stellar density,
which varies along the survey strip.

Since QSO candidates were selected on the basis of their stellar
appearance on photographic plates, low redshift QSOs with detectable
host galaxies on the plate will also be preferentially de-selected
from the final input catalogue.  The visibility of QSO host galaxies
on photographic plates is extremely difficult to model; given the
large variation in host galaxy properties and the large scatter in the
relation between host and nuclear luminosities (Schade, Boyle \&
Letawsky 2000).  However, the  2QZ should be relatively free of
morphological bias against 'extended' QSOs  for $z>0.4$, at which even
the largest host galaxies ($r_{\rm e}\simeq10\,$kpc), will have
angular size of less than 2\,arcsec, and therefore appear unresolved
on the photographic plates with stellar image size typically
2--3\,arcsec.  This is further substantiated by  Hewett, Foltz \&
Chaffee (1995) who find an excess of extended sources in the Large
Bright Quasar Survey (LBQS), only at $z<0.4$.  Also, Meyer et
al. (2001) find that 4 out of 71 QSOs found in the Fornax Cluster
Spectroscopic Survey (FCSS) are extended from APM scans, but that
these are generally mis-classifications due to nearby bright stars and
galaxies.  It should be noted however, that the LBQS and FCSS have
brighter flux limits, $\bj\simeq18.8$ and $\bj\simeq19.7$
respectively, than the 2QZ.  With brighter nuclei, the effect of the
host galaxies in determining the observed morphology will be reduced.

We therefore choose not to derive any completeness correction for the
morphological effect of host galaxies.  For many applications where
the QSO space density is normalized to a local mean (e.g. clustering)
any $z$-dependent  bias (morphological or otherwise) can be corrected
as part of normalization to the observed $n(z)$ relation as long as
that bias is a smooth, slowly varying function of redshift.

\subsection{Photometric completeness}\label{sec:colcomp}

\begin{figure}
\centering 
\centerline{\psfig{file=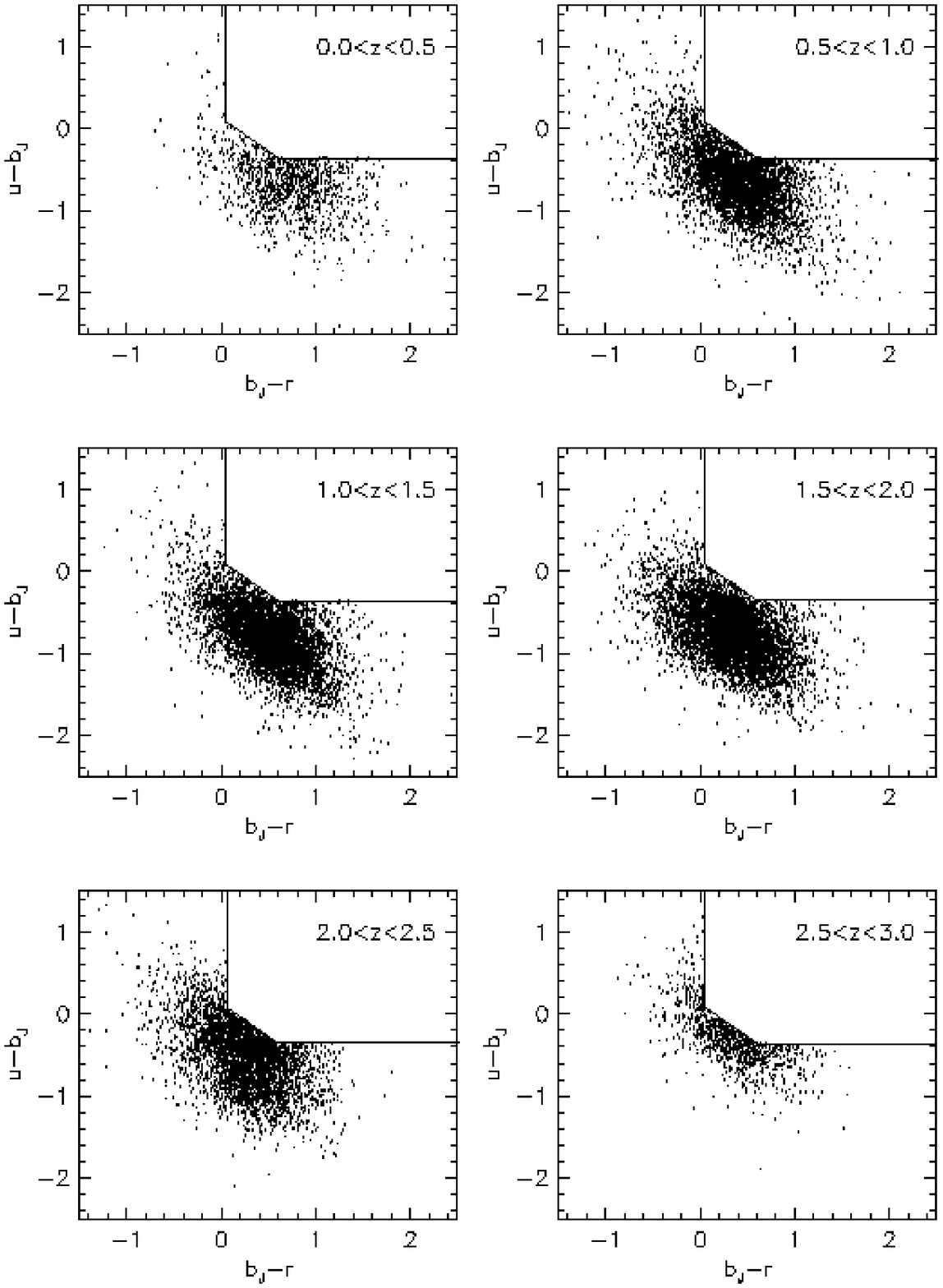,width=8.0cm}}
\caption{Colour-colour plots for 2QZ/6QZ QSOs in redshift bins of
$\Delta z=0.5$.  The solid lines indicate our colour selection
limits.  At low redshift ($0.0<z<0.5$) the QSO colours are redder in
$\bj-r$, therefore the $u-\bj$ cut has greater impact on selection.
At high redshift ($z>2$) the QSO colours become increasingly red,
until the predicted centre of the locus is within the region excluded
by our colour selection limits.}
\label{fig:colcolz}
\end{figure}

Photometric incompleteness is caused by errors in the photographic
magnitudes ($\sim$0.1 mag in each band) and variability (due to the
non-contemporaneous nature of the $u\bj r$ plates on each field) which
cause QSOs to exhibit $u-\bj/\bj-r$ colours outside our selection
criteria.  It could also be the case that we are missing some fraction
of the QSO population that are intrinsically red, however the fraction
of red QSOs in a $\bj$ flux limited sample is likely to be small
\cite{meyer01}.  Hewett, Foltz \& Chaffee (2001) also find that the
fraction of objects missed from the blue selected LBQS is only
$13\pm4$ per cent, using matches to the FIRST radio survey
\cite{first}.  Given the uncertain nature of any red QSO population we
chose to base our photometric completeness estimates on the likelihood
of detecting typical blue QSOs.  In order to fully quantify the
incompleteness due to our colour selection we have adopted a
semi-empirical approach.  This allows us to make use of the measured
properties of the 2QZ QSOs as well as theoretical estimates of QSO
colours.  We derive the mean QSO colours in $u-\bj$ and $\bj-r$ as a
function of $z$ and $\bj$ and use this in combination with the
measured dispersion of 2QZ QSOs about the mean to model the colour
distributions.  Applying our selection criteria to these model QSO
colours allows us to determine the fraction of QSOs that would have
been selected.

\begin{figure*}
\centering 
\centerline{\psfig{file=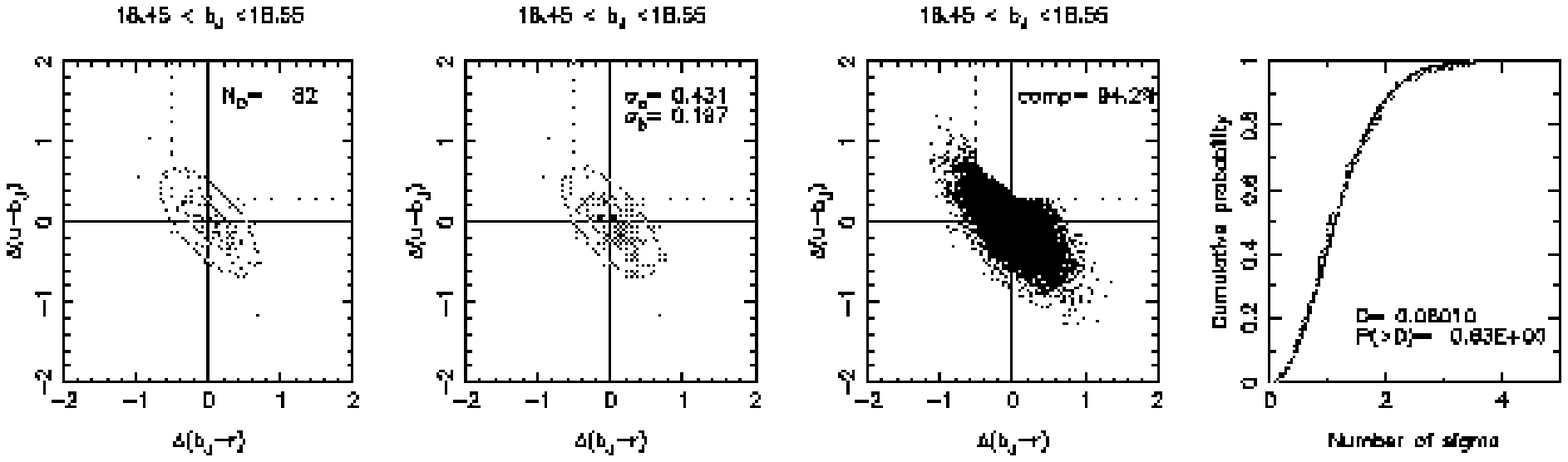,width=16.0cm}}
\vspace{0.5cm}
\centerline{\psfig{file=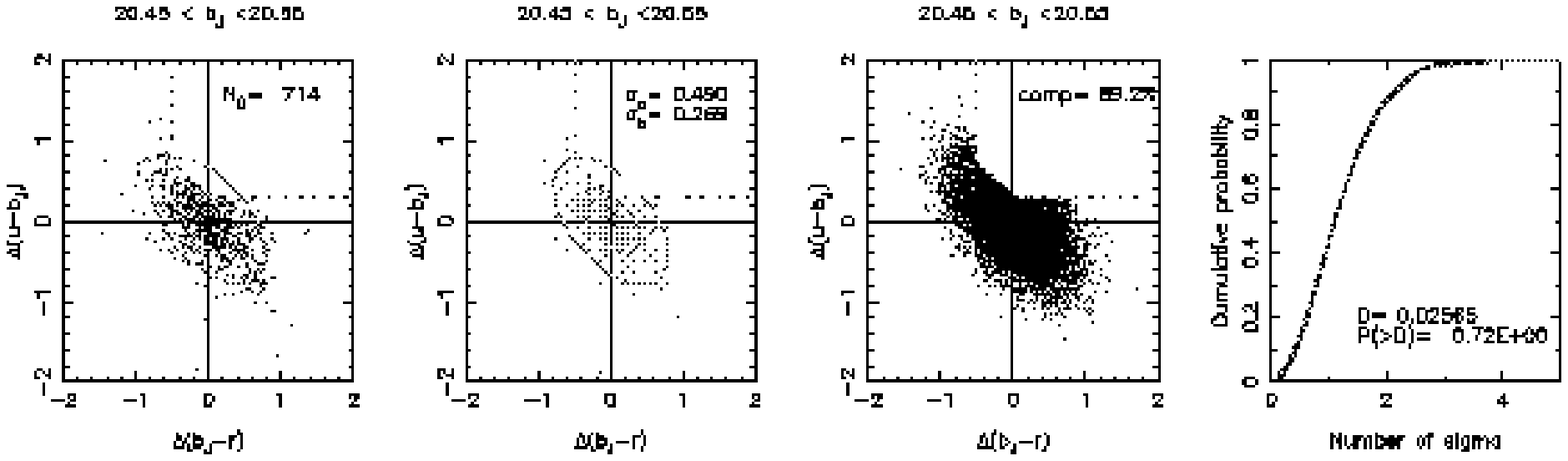,width=16.0cm}}
\caption{Fitting of the $u-\bj$ vs. $\bj-r$ colour distribution for
two example magnitude intervals.  Top: $18.45<\bj<18.55$, Bottom:
$20.45<\bj<20.55$.  From left to right the plots are: a) colour
distributions relative to the mean for QSOs with $1\sigma$ and
$2\sigma$ fitted contours.  $N_{\rm Q}$ is the number of QSOs in the
plot.  b) Grey-scale of the same distribution,
showing the fitted $\sigma$ values.  c) A random realization of the
fitted model with 10000 points and an estimated completeness value. d)
A K-S test between the radial distributions of the QSOs and random
points.}
\label{fig:colfit}
\end{figure*}

Our first step is to derive the mean colour-redshift relations for the
2QZ.  The $u-\bj$ and $\bj-r$ colours of 2QZ/6QZ QSOs are shown in
Fig. \ref{fig:colz} as a function of redshift (small points), also
shown are the mean colours in $\Delta z=0.1$ intervals (connected filled
circles).  The $u-\bj$ vs. $z$ colours show slow variations due to
various emission lines moving in and out of the bands, and an up-turn
at $z=2$ where Lyman-$\alpha$ moves from the $u$-band to the
$\bj$-band, and the $u$-band becomes increasingly affected by
Lyman-$\alpha$ forest absorption.  At low redshift ($z<0.5$) we also
see evidence of a move red-wards in the QSO $u-\bj$ and $\bj-r$
colours.  This is most noticeable in the mean $\bj-r$, as sources with
redder $u-\bj$ colours tend not to be selected.  Indeed, while at most
redshifts there is little evidence of a sharp drop in QSO numbers
redder than the $u-\bj$ cut (dotted line), this is not the case at
$z<0.5$ where most QSOs selected are bluer than the $u-\bj$ cut.  This
effect is most clearly seen in Fig. \ref{fig:colcolz} which displays the
$u-\bj$ vs. $\bj-r$ colours of 2QZ/6QZ QSOs in redshift bins of width
$\Delta z=0.5$.  At $z<0.5$ the QSO locus is more affected by the
$u-\bj$ colour cut, due to the redness of the $\bj-r$ colours.  This
reddening at low redshift is probably due to the increasing amount of
contamination from the host galaxies of the QSOs, confirmed by the
appearance of host galaxy features (\cah\ \& K) in the spectra of low
redshift QSOs in the 2QZ \cite{composite02}.  At high redshift
($z\sim2.5$ and greater) mean QSO colours lay within the region
excluded from our analysis by our colour selection limits, and QSOs
that are detected are found close to the colour selection limits.

We then compared the 2QZ colours to those derived from the SDSS
composite QSO spectrum \cite{sdsscomp}, taking the SDSS composite and
convolving it with the $u$, $\bj$ and $r$ band passes (using those
specified by Warren et al. (1991) and references therein), 
and then zero-pointing the magnitudes to the mean 2QZ colours at
$z=1-2$.  The required shifts are an addition of 0.16 and $-0.08$ mag
to the model $u-\bj$ and $\bj-r$ colours respectively.  We have
examined the uniformity of the 2QZ colours by comparing the median QSO
colours in each UKST field.  The rms scatter found is 0.03 mags, fully
consistent with the minimum errors typically achievable with
photographic plates.  Detailed tests of the procedure used to generate
the synthetic colours found no significant errors.  The exact source
of the above zero-point offsets in unclear, but as the 2QZ photometry
is internally consistent and our completeness estimates are largely
empirical these offsets should not cause any systematic errors in our
analysis of the survey completeness.

We see that the $u-\bj$ and $\bj-r$ SDSS colours (solid lines in
Fig. \ref{fig:colz}) are very close to those of the 2QZ over the
redshift range $z=0.5-2.0$.  At higher redshift, the SDSS colours are
redder than those from the 2QZ, as Lyman-$\alpha$ absorption
progressively reddens the QSO colours.  This demonstrates that the 2QZ
completeness starts to decline past $z=2$ as we preferentially select
only the bluest QSOs at these redshifts.  We also note that the SDSS
colours are bluer in both $u-\bj$ and $\bj-r$ than the 2QZ at $z<0.4$.
We attribute this to the decreased effect of the host galaxy in the
higher luminosity (typically 1 mag) QSOs sampled by the SDSS survey;
based on the flat relation between nuclear and host luminosities for
QSOs ($L_{\rm QSO} \propto L_{\rm gal}^{0.2-0.4}$)
\cite{sbl00,composite02}.  In the SDSS composite this could also be
due to contamination at low redshift by narrow emission line galaxies
\cite{sdsscomp}, as well as the fact that because there are many more
QSOs found at high redshift, the higher redshift QSOs (with less
contribution from the host galaxy) will dominate the construction of
the composite spectrum.

We also compare the 2QZ/6QZ QSO colours to a sample of QSOs which have
been selected on the basis of their variability (Hawkins \& Veron
1995; Hawkins private communication).  We derive the mean $u-\bj$ and
$\bj-r$ colours for Hawkins QSOs with $\bj<21$ as a function of
redshift and plot them in Fig. \ref{fig:colz} (dashed lines).  They
are zero-pointed to the mean 2QZ/6QZ colours at $z=0.7-2.0$.  These
shifts are slightly larger than those derived above for the model
colours based on the SDSS QSO composite and are an addition of $-0.27$
and $-0.16$ mag to the derived $u-\bj$ and $\bj-r$ colours of the
Hawkins sources), and may be due to a zero-point offset in the Hawkins
photometry.  Once this zero-point shift has been accounted for
the mean Hawkins colours follow closely the mean 2QZ/6QZ colours at
$0.5<z<2.3$.  At higher  redshift they become redder than the 2QZ/6QZ,
as do the SDSS colours.  At lower redshift the Hawkins $\bj-r$ colours
are close to the 2QZ/6QZ colours at $z>0.2$, but the Hawkins $u-\bj$
are significantly redder than those of the 2QZ/6QZ.  As the Hawkins
variability selected QSOs are not selected on the basis of colour,
they should, to first order, be independent of colour, and thus follow
more closely the true colour distribution of the QSO population.  We
do note, however, that certain issues can affect variability
selection, such as host galaxy contamination, the correlation of
variability with luminosity and redshift (e.g. Tr\`evese \& Vagnetti
2002), and these may introduce some colour dependence to any
variability selected sample.

The above considerations suggest that we use the following to define
our mean QSO colour-redshift relations: i) the mean Hawkins colours
for $u-\bj$ at $z<0.5$; ii) the mean 2QZ/6QZ colours for $u-\bj$ at
$0.5<z<2.0$ and $\bj-r$ at $0.0<z<2.0$; iii) the SDSS composite
colours at $z>2.0$ for $u-\bj$ and $\bj-r$.  We note that in each case
the SDSS composite and Hawkins colours have had the above zero-point
shifts applied.

A final point to consider is whether there is any magnitude dependence
of the mean colours.  In particular, if the reddening of the $\bj-r$
colours at low $z$ is due to increasing host galaxy contamination,
this effect might be most noticeable in intrinsically faint QSOs.  We
find this to be the case for 2QZ/6QZ $\bj-r$ colours.  The
variation in colour can be described by
\begin{equation}
\Delta(\bj-r)=-9.17+0.479\bj+13.1z-0.685\bj z
\label{eq:deltabr}
\end{equation}
at $z<0.70$.  At higher redshift no significant effect is seen.  We
see no similar reddening of the $u-\bj$ colours towards fainter
magnitudes for low redshift QSOs in the 2QZ/6QZ, although some weak
dependence would be expected if it is the host galaxy which is
producing the variation in $\bj-r$ colours.  We have tested the
expected $\Delta(u-\bj)$ for a given $\Delta(\bj-r)$ by combining the
SDSS QSO composite spectrum and model galaxy spectra.  For a range of
galaxy SEDs (e.g. single instantaneous bursts of star formation of age
between 3 and 10 Gyr) the typical $\Delta(u-\bj)$ is between $\sim0.2$
and 0.5 times $\Delta(\bj-r)$ above $z\sim0.2$.  As the exact nature
of this correction is difficult to model due to uncertainties in the
galaxy SED and the fractional contribution of the host, and we do
not see any evidence of a magnitude dependence of the mean $u-\bj$
colours, we have decided not to make any correction to the mean
$u-\bj$ as a function of QSO magnitude.  We therefore correct our mean
colour-redshift in $\bj-r$ only by an amount defined from
Eq. \ref{eq:deltabr}, although we limit the $\bj-r$ colours to be no
bluer than those from the SDSS composite.

\begin{figure}
\vspace{-0.5cm}
\centerline{\psfig{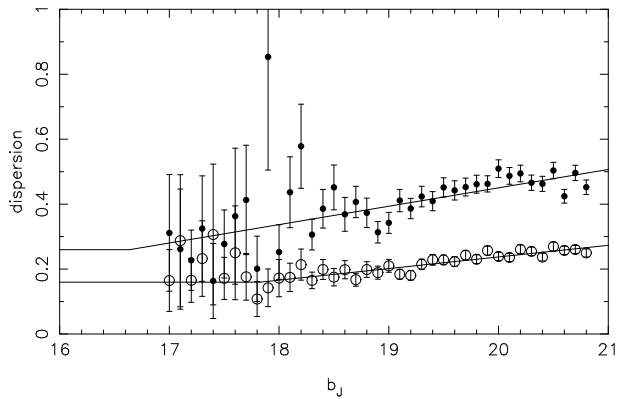}}
\caption{The semi-major axis (filled circles) and semi-minor axis (open
circles)  dispersions for the 2D Gaussian distribution fitted to the
colour-colour distributions as a function of magnitude.  The solid
lines denote the best fit lines used to derive the model QSO colour
distributions.  At bright magnitudes the dispersions are forced to be
no less than the mean dispersion found at $\bj=17-18$.} 
\label{fig:colsig}
\end{figure}

\begin{figure}
\centerline{\psfig{file=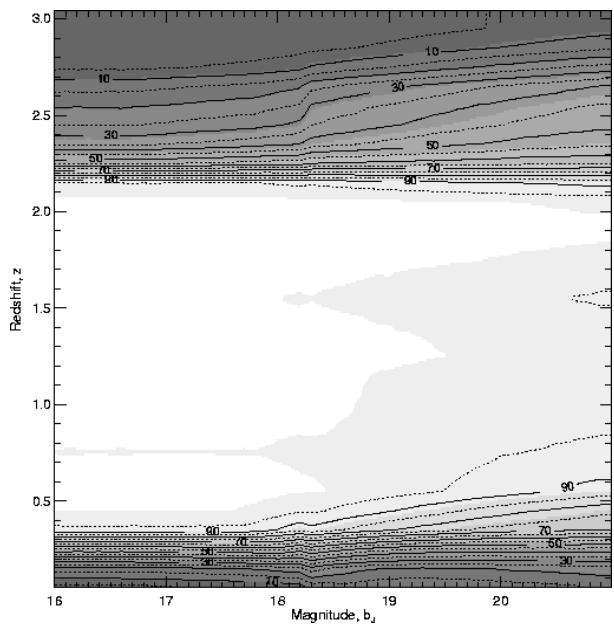,width=8.0cm}}
\caption{Photometric completeness contours in the ($\bj, z$) plane for 
the 2QZ.  The contours are plotted at 5 per cent intervals: 
dotted lines denote 5, 15, 25,...95 per cent completeness; solid 
lines correspond to 10, 20, 30...90 per cent.}
\label{fig:colcomp}
\end{figure}

Once the mean colour-redshift relations are defined, we then determine
the dispersion about the mean.  In doing this we assumed that the
dispersion was a function of magnitude only, a reasonable assumption
given that photometric errors and variability dominate the scatter.
Any redshift dependence of variability was considered to be a second
order effect.  The dispersion about the mean $u-\bj$ and $\bj-r$
colours was fit by a 2D Gaussian distribution in $\Delta\bj=0.1$
intervals between $z=1$ and 2.  This redshift range was chosen as the
regime where completeness was highest, so as to give the most accurate
description of the QSO colour distribution.  Two examples of these
fits are shown in Fig. \ref{fig:colfit}.  Regions outside of our colour
selection limits were not included in the fitting procedure.  We first
attempted to fit the colour distribution with four parameters: peak
height, semi-major axis ($\sigma_{\rm a}$), semi-minor axis
($\sigma_{\rm b}$) and position angle.  However investigation showed
that in all cases the position angle was close to 45$^{\circ}$.  The
fitting process was then repeated fixing the position angle to be
45$^{\circ}$.  Random distributions following the fitted Gaussian were
produced to confirm that the 2D Gaussian was a reasonable fit to the
data via a radial  Kolmogorov-Smirnov (K-S) test.  Only two
$\Delta\bj=0.1$ intervals showed a significant difference between the
data and the model.  In general we note that the model does generate
marginally less extreme outliers in the colour distributions, however
this does not significantly affect the determined completeness values.
The measured widths of the 2D Gaussian model were fitted by a straight
line in the range $17<\bj<20.85$, and these fits used to determine the
dispersion in our final Monte-Carlo simulations that estimate QSO
completeness.  At $\bj<17$ there are insufficient QSOs to make an
adequate fit to their distribution, so the dispersions are fixed to be
the mean of the dispersion in the $\bj=17-18$ range (see
Fig. \ref{fig:colsig}).

Finally we make Monte-Carlo simulations of the QSO colour
distributions in $\Delta\bj=0.1$ and $\Delta z=0.1$ bins using 100000
QSOs in each interval.  The fraction of simulated QSOs which fall
outside our colour selection limits then define our incompleteness.
The final photometric completeness contours are shown in
Fig. \ref{fig:colcomp}.  The photometric completeness is largely
independent of magnitude and is at least 70 per cent or greater over
the redshift range $0.4<z<2.2$.  At higher redshifts the completeness
rapidly drops, falling to below 50 per cent at $z > 2.3$.  The bluer
$u-\bj$ limit imposed on the 6QZ ($u-\bj<-0.5$ rather than
$u-\bj<-0.36$) causes a weak step in the completeness contours which
can be seen at $\bj=18.25$.

\subsection{Coverage completeness}

\begin{figure}
\centering
\centerline{\psfig{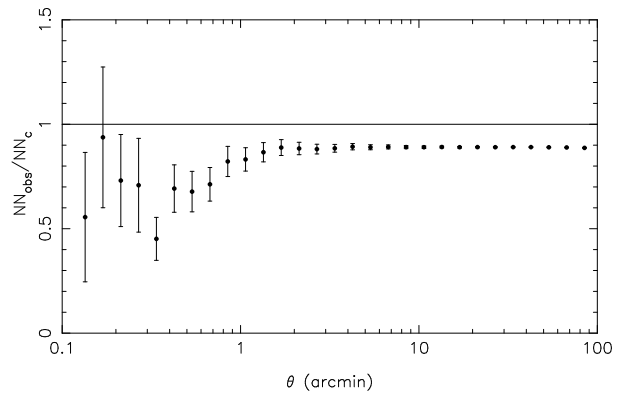}}
\caption{The ratio of close pairs found in the 2QZ input catalogue,
$NN_{\rm c}$ to those observed with 2dF $NN_{\rm obs}$.  At scales
less that $\sim1-2$ arcmin there is a decrease in the number of pairs
due to the positioning constrains of the 2dF fibre positioner.  The
constant offset below 1 at larger angular scales is cause by the lower
mean density of observed objects compared to the full catalogue
(i.e. not all objects could be observed).}
\label{fig:closepairs}
\end{figure}

\begin{figure*}
\centering
\centerline{\psfig{file=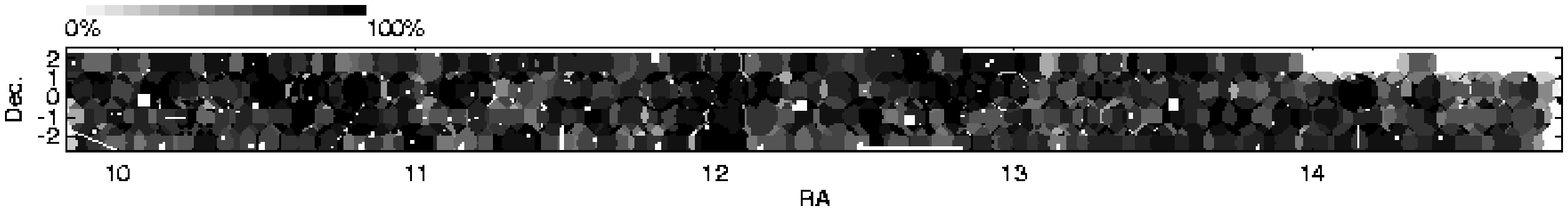,width=18cm,angle=90}}
\centerline{\psfig{file=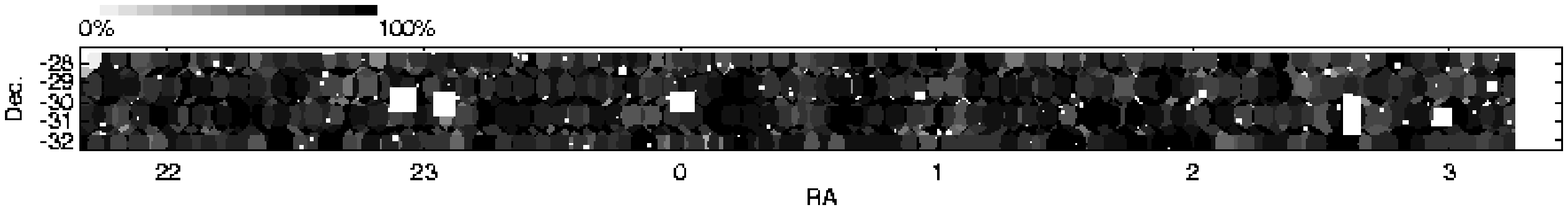,width=18cm,angle=90}}
\caption{The completeness map of the 2QZ catalogue for the equatorial
(top) and SGP (bottom) regions.  The grey-scale indicates the
percentage of 2QZ candidates observed and positively identified
(quality 1) over the two survey strips.  This is equivalent to
$\fc(\theta)\fs(\theta)$.}
\label{fig:comp}
\end{figure*}

Of all four types of completeness, at a sector level the coverage
completeness (or coverage) is the simplest to derive, being only a
function of angular position.  We define the coverage as being the
ratio of observed to unobserved sources in each of the sectors defined
by overlapping 2QZ fields, we denote this by $\fc(\theta)$.
The coverage maps (pixelized on 1 arcminute scales) of the two 2QZ
strips are shown in Fig. \ref{fig:cover}.  These maps also account for
holes in the survey regions which are due to bright stars or
photographic plate defects/features.  The incompleteness in coverage
arises from the difficulty of configuring all targets onto 2dF fibres
using the adaptive tiling scheme \cite{2dfgrs}.  This is caused by
variations in target density, a small (typically $\sim1$ per cent)
number of broken or unusable 2dF fibres, and constraints due to the
minimum fibre spacing ($\sim30$ arcsec).  These issues are in large
part alleviated by the large overlaps between 2dF fields, however it
was still not possible to observe 100 per cent of the QSO candidates
in the 2QZ.  The remaining decline in the number of close pairs is
demonstrated in Fig. \ref{fig:closepairs} which shows the ratio of
close pairs from observed ($NN_{\rm obs}$) to those in the input
catalogue ($NN_{\rm c}$).  This shows a decline in close pairs at
angular scales less than $\sim1-2$ arcmin.  This has no significant
effect on the clustering analysis of QSOs (e.g. see Croom et al.
2001a).  The number of close pairs contained within the input
catalogue is also limited by the resolution available on the APM scans
of our UKST plate material.  This reduces the number of close pairs on
angular scales less than $\sim8$ arcsec \cite{2qzpaper3}.    

\subsection{Spectroscopic completeness} 

\subsubsection{Position dependent spectroscopic completeness,
  $\fs(\theta)$}

Spectroscopic completeness is perhaps the most complex of all
completenesses to describe.  In our discussion below we focus on the
spectroscopic completeness of the QSOs in our survey, as they are the
primary targets of our observations.  In general different classes of
object will have different spectroscopic completeness functions,
e.g. NELGs are more easily identified in low $S/N$ spectra than
stars.  Our assessment of QSO spectroscopic incompleteness 
assumes that there are equal fractions of QSOs among both the
identified and unidentified spectra (see Section \ref{sec:2qzcat}).
Alternative methods may be required to assess the completeness of
different object classes.

  The distribution of combined coverage and spectroscopic
incompleteness, $\fc(\theta)\fs(\theta)$,  across the two 2QZ strips
based on quality 1 IDs alone is shown in Fig. \ref{fig:comp}
(c.f. Fig. \ref{fig:cover} which shows the coverage incompleteness
only).  However, this only describes the mean spectroscopic
completeness in each sector (i.e. as a function of angular position).
This is straightforward to calculate on a sector-by-sector basis,
based on the ratio $N_1/N_{\rm obs}$, where $N_1$ is the number of
quality 1 IDs in a sector and $N_{\rm obs}$ is the number of targets
observed in a sector.  By contrast, $\fc$ is determined by the ratio
$N_{\rm obs}/N_{\rm c}$, where $N_{\rm c}$ is the total number of 2QZ
sources within the sector.  However, while coverage is only dependent
on angular position, spectroscopic completeness will in general also
be dependent on $\bj$ and $z$.  For many analysis (e.g. the luminosity
function, see below), we only require global magnitude and redshift
dependent corrections.  These will vary with the minimum sector
completeness level applied to produce any `complete' spectroscopic
catalogue (see Section 2.1).  For example the inclusion of more low
mean completeness sectors will produce a spectroscopic completeness
function with a more marked dependence on $\bj$.

\begin{figure}
\centering 
\centerline{\psfig{file=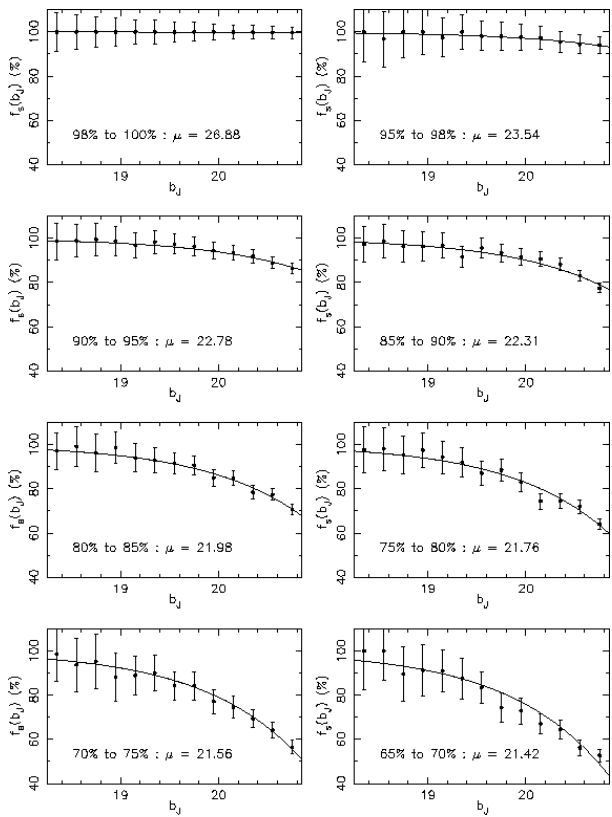,width=8cm}}
\caption{Spectroscopic completeness as a function of $\bj$ magnitude,
$\fs(\bj)$, for sectors with different mean completeness values,
$\fs(\theta)$  (filled circles).  These are plotted for
mean sector completeness intervals 0.98--1.00, 0.95--0.98, 0.90--0.95,
0.85--0.90, 0.80--0.85, 0.75--0.80, 0.70-0.75 and 0.65--0.70.  In each
case the best fit magnitude dependent completeness model is plotted
(solid line), and the best fit value of $\mu$ is shown.}
\label{fig:spcom}
\end{figure}

\subsubsection{Magnitude dependent spectroscopic completeness, $\fs(\bj)$}

\begin{figure}
\centering 
\centerline{\psfig{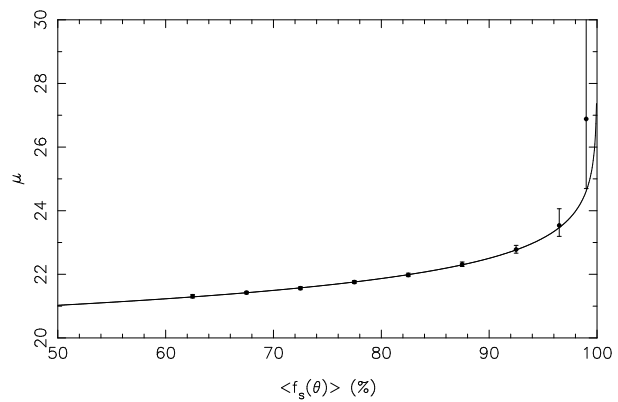}}
\caption{The magnitude dependent spectroscopic completeness $\mu$
parameter as a function of mean sector spectroscopic completeness,
$<\fs(\theta)>$ (solid points).  These are the best fit
parameters for the fits shown in Fig. \ref{fig:spcom}.  The solid line
shows the function used to describe the dependence of $\mu$ on
$<\fs(\theta)>$.}
\label{fig:mufit}
\end{figure}

Although for many purposes a global magnitude dependent completeness
correction would suffice, it is also possible to determine the
magnitude dependence of spectroscopic completeness on a
sector-by-sector basis.  In Fig. \ref{fig:spcom} we plot the
spectroscopic completeness as a function of $\bj$ for sectors with
different mean spectroscopic completeness levels.  In sectors of high
completeness there is very little dependence of completeness on $\bj$,
while for less complete sectors (typically observed in worse
conditions) the completeness has a stronger dependence on $\bj$.
Following a similar approach to Colless et al. (2001) we parameterize
this magnitude dependence as a function of the form
\begin{equation}
\fs(\bj)=1-\exp(\bj-\mu),
\label{eq:fsbj}
\end{equation}
where $\mu$ is the single free parameter.  Fig. \ref{fig:spcom} also
shows the best fit function for each mean completeness interval.  This
simple functional form traces well the dependence of completeness on
$\bj$ magnitude.  We can then relate the best fit $\mu$ values to the
mean completeness  $<\fs(\theta)>$.  This is plotted in
Fig. \ref{fig:mufit}, and can be fit by the function
\begin{equation}
\mu=A+B\ln(1-<\fs(\theta)>),
\end{equation}
where $A$ and $B$ are constants to be fit for.  Their best fit
values are $20.388\pm0.77$ and $-0.919\pm0.052$ respectively.  The
magnitude dependence of $\fs$ is then fully described by
\begin{equation}
\fs(\bj)=1-\frac{\exp(\bj-A)}{[1-\fs(\theta)]^B}.
\end{equation}
We are then able to combine this with the mean spectroscopic
completeness within a sector, $\fs(\theta)$, to derive the
full position and magnitude dependent spectroscopic completeness
correction, 
\begin{equation}
\fs(\theta,\bj)=\frac{N_{\rm obs}(\theta)}{N_{\rm
 est}(\theta)}\fs(\theta)\fs(\bj).
\end{equation}
The value $N_{\rm est}(\theta)$ is an estimate of the number of
sources with quality 1 IDs given the function $\fs(\bj)$ for a
particular sector, and is given by
\begin{equation}
N_{\rm est}(\theta)=\sum^{N_{\rm
obs}(\theta)}_{i=1} \fs(\bj).
\end{equation}
Therefore in order to derive the positional and magnitude dependence
of the spectroscopic completeness of the 2QZ we require the mean $\fs$
value in each sector, $\fs(\theta)$, and the derived $N_{\rm
 est}(\theta)$ value for each sector.  These are provided as
part of the 2QZ data release.

\begin{figure}
\centering 
\centerline{\psfig{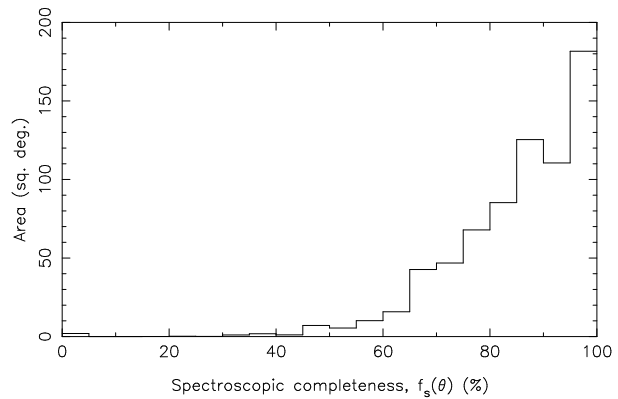}}
\caption{The area of the 2QZ as a function of mean sector
spectroscopic completeness, $\fs(\theta)$.}
\label{fig:spcomparea}
\end{figure}

The simple functional form for $\fs(\bj)$ in Eq. \ref{eq:fsbj}  can be
fit to the full 2QZ catalogue, and this results in
$\mu=22.168\pm0.026$.  Given the relation between $\fs(\theta)$
and $\mu$ derived above, the overall spectroscopic incompleteness of
85.7 per cent would imply a value of $\mu=22.18$, in excellent agreement with
the directly fitted value.  The 2QZ catalogue limited to sectors with
at least 70 per cent quality 1 identifications gives
$\mu=22.407\pm0.037$.  This 70 per cent catalogue, based on
Table \ref{tab:numbers}, provides a suitable compromise between the
need to achieve relatively high level of mean overall spectroscopic
completeness (89 per cent), whilst retaining as many of the original
quality 1 QSOs in the final catalogue (20905) as possible (see
Fig. \ref{fig:spcomparea}).

\begin{figure}
\centering 
\centerline{\psfig{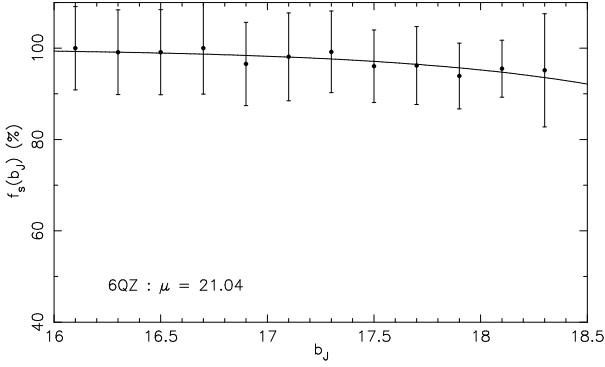}}
\caption{Spectroscopic completeness as a function of $\bj$ magnitude,
$\fs(\bj)$, for the 6QZ (filled circles).  Also shown is the best
fitting model to the magnitude dependent completeness (solid line).}
\label{fig:spcom6qz}
\end{figure}

The 6QZ sample has a mean spectroscopic completeness of 97.1 per cent,
considerably higher than that of the 2QZ.  This mean correction should
be sufficient to account for spectroscopic incompleteness, as there is
only weak (unsignificant) magnitude dependence on completeness in the
6QZ.  However, if we do apply a similar analysis to that described
above, and fit Eq. \ref{eq:fsbj} to the 6QZ over the range
$16<\bj<18.25$ we find $\mu=21.04^{+1.35}_{-0.55}$ (see Fig
\ref{fig:spcom6qz}).  This can also be used to describe magnitude
dependent spectroscopic completeness in the 6QZ.  Note that the
relation between $\mu$ and $\fs(\theta)$ derived above for the 2QZ is
not valid for the 6QZ.

\subsubsection{Redshift dependent spectroscopic completeness, $\fs(z)$}

There is also the possibility that the spectroscopic identification
procedure has also introduced subtle redshift-dependent biases.  This
can happen in low $S/N$ spectra, particularly when only one strong QSO
emission line is visible in the observed spectral window.  On
inspection of the smooth 2QZ $n(z)$ relation (Fig. \ref{fig:nzcat}),
we infer that any sudden changes in the efficiency of redshift
determination must be small, at least over the
$0.4<z<2.2$ redshift range, where the photometric completeness is
also high ($>70$ per cent).  The redshifts at which strong emission
lines move into the observed frame (e.g. \ciii\ $\lambda1909$  at
$z=1.00$, \civ\ at $z=1.45$ or Ly$\alpha$ at $z=2.12$) are not marked
by a significant increase in the $n(z)$.  At $z>0.35$ there is a
significant discontinuity in the $n(z)$, most likely caused by the
appearance of \mgii\ emission line in the observed spectral window.  At
these redshifts, in many cases we note that objects classified as a
QSO by the appearance of a broad \mgii\ emission line often do not
exhibit significant visible broad H$\beta$ or H$\gamma$ emission
lines.  We conclude, therefore, that there is likely to be a significant
number of objects at $z<0.35$ which we have failed to classify as QSOs
based only on the appearance of the Balmer series lines.  

In order to perform quantitative tests on the redshift dependence of
spectroscopic completeness we carry out K-S tests
comparing the redshift distributions of QSOs selected in sectors of
different completeness.  We first compare QSOs from sectors of
60--70\%, 70--80\%, 80--90\% and 90--100\% to the $n(z)$ distribution
of all QSOs.  The derived probabilities of the null hypothesis that
the $n(z)$ distributions are the same are 0.318, 0.193, 0.853 and
0.133 respectively.  There is therefore no significant ($>1.5\sigma$)
difference between these $n(z)$ distributions.  We secondly compare the
$n(z)$ distributions for the lower completeness QSO samples, 60--70\%,
70--80\% and 80--90\%, to the distribution for QSOs with 90--100\%
completeness.  The resulting probabilities of the null hypothesis in
this case are 0.182, 0.007 and 0.040 respectively.  This does
therefore indicate some small but significant differences.  The lower
completeness samples are generally biased such that they have slightly
higher numbers of high redshift QSOs.  This is suggestive that the
greater number of strong emission lines at high redshift is making it
easier to identify lower $S/N$ QSO spectra.

\begin{figure}
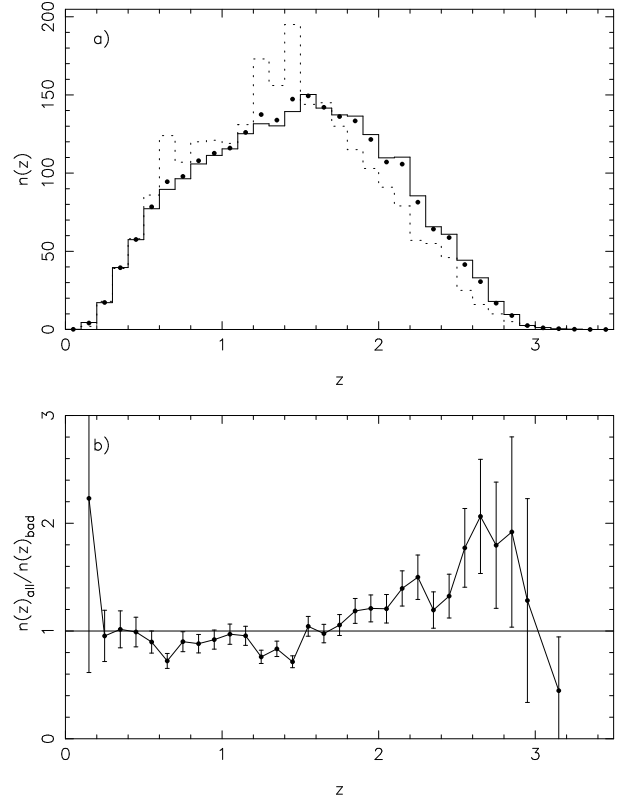

\centering
\centerline{\psfig{file=nz_all_bad.ps,width=8cm}}
\vspace{0.3cm}
\centerline{\psfig{file=nz_ratio.ps,width=8cm}}
\caption{The redshift dependence of spectroscopic completeness. a) The
observed redshift distribution of all 2QZ QSOs, $n(z)_{\rm all}$ (solid
line), compared to the redshift distribution of objects with low quality
identifications which were subsequently re-observed and given a high
quality QSO redshift, $n(z)_{\rm bad}$ (dotted line).  $n(z)_{\rm
all}$ is renormalized to contain the same total number of QSOs as
$n(z)_{\rm bad}$.  We also show the predicted $n(z)$ assuming that the
incompleteness in the 2QZ is distributed as $n(z)_{\rm bad}$ (filled
points).  b) The ratio of $n(z)_{\rm all}$ to $n(z)_{\rm
bad}$ (renormalized) as a function of redshift.  Error bars are
Poisson.}
\label{fig:speccompz}
\end{figure}

Finally we attempt to derive the completeness corrections required to
correct for these small variations in completeness as a function of
redshift.  If there were no systematic variations of completeness with
redshift, the total fractional spectroscopic completeness, $\fs$,
would simply be $\fs\equiv\fs(\theta,\bj)$.  However, this will be
modulated by any redshift dependent spectroscopic incompleteness.  In
order to determine this modulation of the redshift distribution we
turn to the repeat observations.  There are 2473 QSOs which have one
spectrum of quality 11 and a second of lower quality (quality 22 or
worse, including objects with ?? IDs).  We can then use these objects
to look at the {\it redshift distribution of unidentified objects}.
In Fig. \ref{fig:speccompz}a we plot the redshift distribution of all
quality 11 QSOs, ($n(z)_{\rm all}$, solid line) and the redshift
distribution for those objects which had both low and high quality
identifications in repeat observations ($n(z)_{\rm bad}$, dotted
line).  In both cases the redshift used is that from the best quality
ID (quality 11) which is treated as the ``true'' redshift.  There
is a clear and significant difference between the two (note that
$n(z)_{\rm all}$ has been renormalized).  We see that a larger
fraction of bad IDs are present at $z\sim0.6$ and $z\sim1.4$, while
there are fewer bad IDs at $z>1.8$.  These appear to be due to changes
in the visibility of strong emission lines as a function of redshift.
The ratio of $n(z)_{\rm all}$ to $n(z)_{\rm bad}$ is shown in
Fig. \ref{fig:speccompz}b and more clearly demonstrates the dependence
of completeness on redshift.  We then determine the relative effect
that incompleteness of this type would have on the observed redshift
distribution of the QSOs.  In the full 2QZ sample 14.2\% of sources
have quality poorer than quality class 1.  If this fraction is
distributed in redshift as $n(z)_{\rm bad}$ and added onto the
observed $n(z)_{\rm all}$ we obtain the filled points in
Fig. \ref{fig:speccompz}a (also assuming that the fraction of QSOs in
the unidentified sources is equal to the fraction of QSOs in the
identified sources, see Section \ref{sec:2qzcat}).  We see that making
the correction changes only slightly the overall shape of the derived
QSO $n(z)$.

In order to make the above corrections to $\fs$, we define a function
$R(z)$ such that
\begin{equation}
R(z)=\frac{n(z)_{\rm all}}{n(z)_{\rm bad}},
\end{equation}
where both $n(z)_{\rm all}$ and $n(z)_{\rm bad}$ are normalized such
that the sum of all redshift bins is equal to one.  This ratio is
tabulated in Table \ref{tab:fsz}.  The completeness is a given
sector and $\bj$ interval is then distributed according to this
relation such that
\begin{equation}
\fs(\theta,\bj,z)=\left\{1+\frac{1}{R(z)}\left[\frac{1}{\fs(\theta,\bj)}-1\right]\right\}^{-1}.
\end{equation}  
This relation can then be used to describe the variations in
completeness in a general sense for the 2QZ.  We note that
several assumptions have been made in our estimate of redshift
spectroscopic incompleteness.  In particular, we assume that redshift
dependence is separable from $\fs(\theta,\bj)$.  That is, the
form of $R(z)$ does not vary depending on sector or $\bj$ magnitude.    

\begin{table}
\centering
\caption{Derived values for $R(z)$ (and errors, $\sigma_{R(z)}$)
that can be used to estimate the redshift dependence of spectroscopic
incompleteness.  Data are binned in $\Delta z=0.1$ bins.}
\label{tab:fsz}
\begin{tabular}{rrrrrr}
\hline
$z$ & $R(z)$ & $\sigma_{R(z)}$ & $z$ & $R(z)$ & $\sigma_{R(z)}$\\ 
\hline 
 0.15 &  2.232 &  1.617 & 1.65 &  0.977 &  0.086\\
 0.25 &  0.955 &  0.238 & 1.75 &  1.056 &  0.097\\
 0.35 &  1.016 &  0.171 & 1.85 &  1.187 &  0.116\\
 0.45 &  0.991 &  0.137 & 1.95 &  1.210 &  0.125\\
 0.55 &  0.898 &  0.103 & 2.05 &  1.207 &  0.132\\
 0.65 &  0.723 &  0.070 & 2.15 &  1.395 &  0.163\\
 0.75 &  0.901 &  0.092 & 2.25 &  1.500 &  0.206\\
 0.85 &  0.882 &  0.085 & 2.35 &  1.195 &  0.168\\
 0.95 &  0.920 &  0.089 & 2.45 &  1.324 &  0.203\\
 1.05 &  0.970 &  0.094 & 2.55 &  1.772 &  0.365\\
 1.15 &  0.956 &  0.088 & 2.65 &  2.064 &  0.530\\
 1.25 &  0.760 &  0.062 & 2.75 &  1.796 &  0.585\\
 1.35 &  0.835 &  0.071 & 2.85 &  1.919 &  0.883\\
 1.45 &  0.715 &  0.055 & 2.95 &  1.283 &  0.946\\
 1.55 &  1.044 &  0.092 & 3.15 &  0.446 &  0.499\\
\hline
\end{tabular}
\end{table} 

\begin{figure}
\centering
\centerline{\psfig{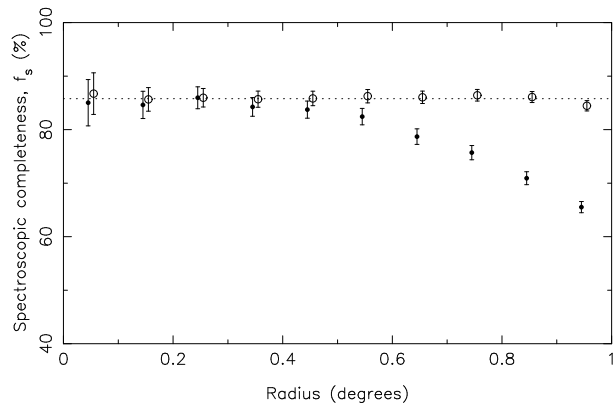}}
\caption{The observed spectroscopic completeness within 2dF fields as
a function of radius from the field centre.  We plot the completeness
for specific observations in a given field (filled circles), and the
completeness for the best observation of a given object including
overlapping fields (open circles).  Also plotted is the mean
completeness from the best observations (dotted line).}
\label{fig:2dffldcomp}
\end{figure}

\subsection{Completeness within 2dF fields}

All the above analysis of completeness is based on the assumption that
within a single 2dF field, or more exactly, a single sector, the
survey completeness is uniform.  However, various effects could cause
systematic variations within these regions.  E.g.  Croom et al. (2001)
showed that there is a systematic variation in the effective
throughput across a single 2dF field (see their Fig. 2).  After
further investigation of this effect, it was found to be due to
systematic residuals remaining from sky subtraction, caused by the
spectral PSF of the spectrograph degrading slightly near the edge of
the CCDs (e.g. Willis, Hewett \& Warren 2001).  This was due to the
{\small 2DFDR} software effectively basing the throughput estimate of
fibres on the peak flux in sky lines rather than the total flux in the
lines.  As a result, routines to measure fibre throughput from total
sky line flux were written and incorporated into the {\small 2DFDR}
pipeline.  These routines removed the majority of the residual sky
emission, however it should be noted that some excess sky residuals
will still remain due to subtraction of a mean sky spectrum which does
not have exactly the same spectral PSF as the data (see Willis et
al. 2001).  Also, the revised version of {\small 2DFDR} could not be
run on data taken before July 2001, so that in the final 2QZ catalogue
completeness variations as a function of fibre number remain.  As any
given fibre can be positioned over a large fraction of the 2dF
field-of-view this incompleteness as a function of fibre number does
not have a significant effect on the spatial distribution of
completeness, and does not impact on analysis such as the study of QSO
clustering.

\begin{figure}
\centering
\centerline{\psfig{file=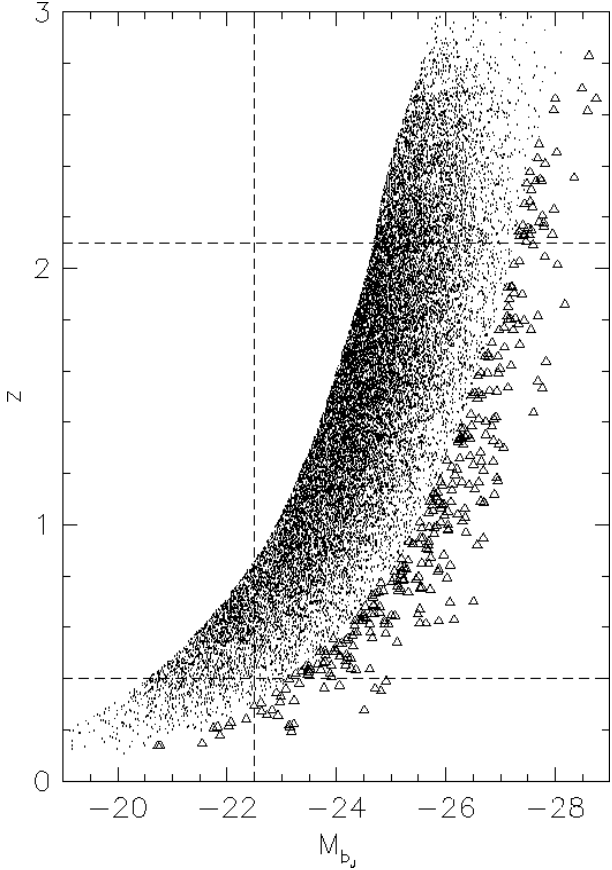,width=8cm}}
\caption{($\mb, z$) distribution for 2QZ (points) and 6QZ (triangles)
QSOs.  Dashed lines indicate the limits of our maximum likelihood
analysis, $\mb<-22.5$ and $0.4<z<2.1$.  For this plot we assume
$\om=0.3$, $\ol=0.7$ and $\ho=70$\kmsmpc.}
\label{fig:zmb}
\end{figure}

Due to a number of different effects, the completeness near the edge
of a 2dF field could be lower than in the central region.
We therefore look at the completeness within a 2dF field as a function
of radius from the field centre.  We first take each individual
observation of a 2dF field and find the spectroscopic completeness as
a function of radius from the field centre.  In
Fig. \ref{fig:2dffldcomp} we plot the number weighted average of these
completeness estimates (filled circles).  A decline in completeness
with increasing radius is visible at $>0.4^{\circ}$, with regions at
$>0.9^{\circ}$ on average only $\sim65$ per cent complete.  This could
be caused by a number of effects.  Systematic errors in astrometry or
field rotation would tend to be worse towards the edge of the 2dF
field.  Similarly, atmospheric refraction effects, if a field was
observed at a different hour angle than it was configured for, would also
be most noticeable at the edges of the field.  In order to see how this
radially dependent completeness affects the final 2QZ catalogue, we
then derive the radially dependent completeness, using only the best
observation of each object, and include objects that are within a
given field, but actually observed in an overlapping field.  This
radial completeness estimate is also shown in
Fig. \ref{fig:2dffldcomp} (open circles).  When using only the best
observations and including overlaps, we find that there is no
significant decline in the completeness at large radii.  This suggests
that the final 2QZ catalogue should not be seriously affected by
radially variable incompleteness.  The most extreme test of the
catalogue uniformity is in clustering analysis.  We have performed
detailed simulations of the radial completeness variations and their
effect on the two-point correlation function, and confirm that these
radial variations have negligible impact on the clustering analysis
(Croom et al. 2003 in preparation).  These tests also imply that any
variation in completeness as a function of fibre number do not effect
measurements of clustering.
  
\section{The optical QSO luminosity function}\label{sec:lf}

\subsection{Data}

\begin{figure*}
\centering 
\centerline{\psfig{file=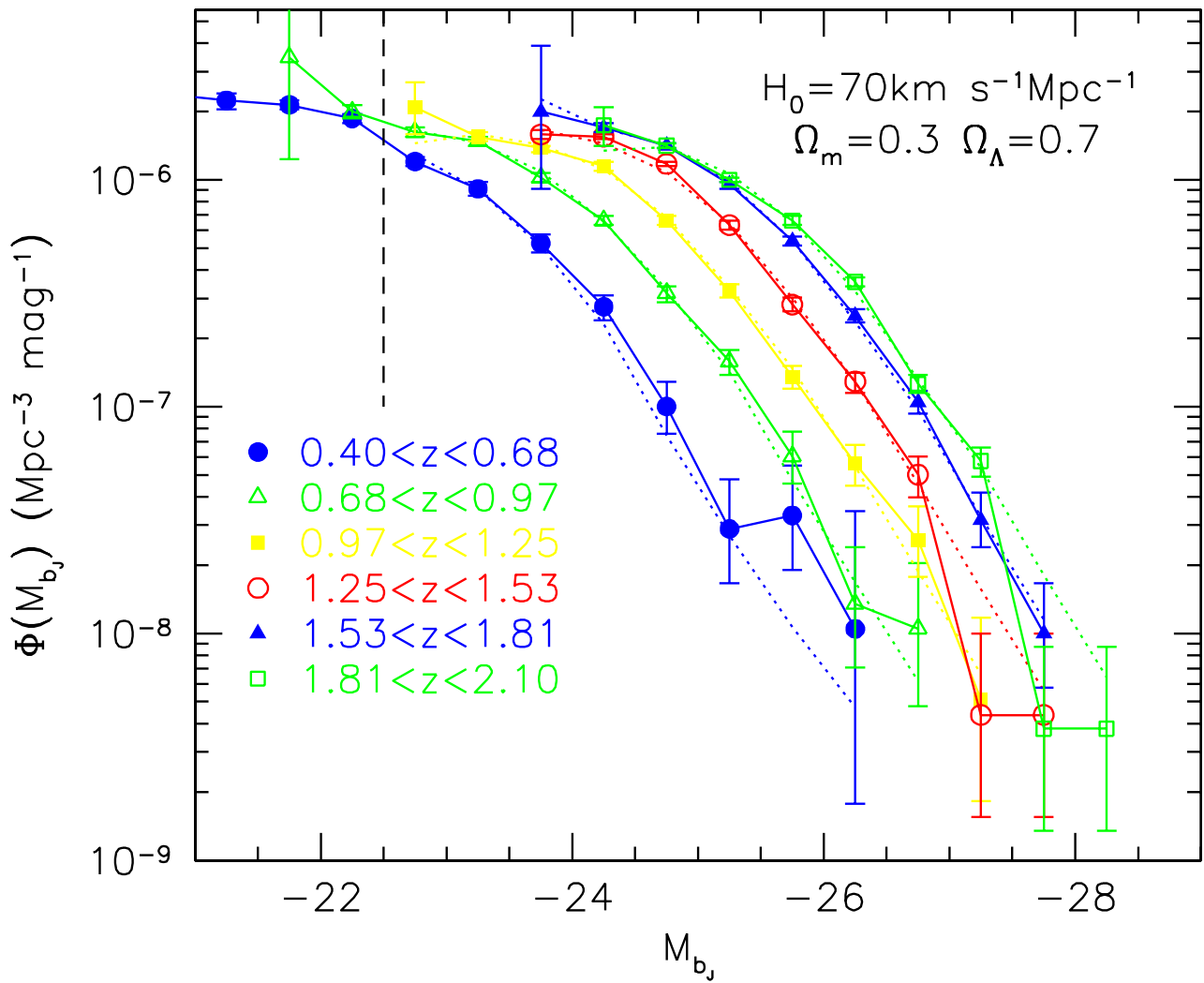,width=18cm}}
\vspace{-2.0cm}
\caption{The optical QSO luminosity function for the 2QZ + 6QZ sample
in six redshift intervals (open and filled points).  The dotted lines
denote the predictions of the  best fit  double power law exponential
evolution model to dataset with $\mb <-22.5$ in the $\Lambda$
Universe.  The model values are calculated based on the accessible
volume in each bin, exactly as the data, and so contain an explicit
correction for incomplete bins.  This means that the dotted lines may
not appear to follow an exact power law.  The vertical dashed line
indicates the $\mb<-22.5$ limit used in the maximum likelihood
fitting.}
\label{fig:olf} 
\end{figure*}
 
Based on the discussion above, we chose to use the 2QZ sample defined
by a minimum mean sector spectroscopic completeness of 70 per cent,
and the full SGP declination strip for the 6QZ.  The effective survey
area corresponds to 595.9 deg$^{2}$ for the 2QZ and 333.0 deg$^{2}$
for the 6QZ after coverage correction.  All QSO magnitudes and sector
magnitude limits were corrected for galactic extinction on the basis
of the mean extinction values \cite{sfd98} in each sector.

A strength of the combined 2QZ/6QZ data set is its homogeneous
selection and photometry.  With the statistical precision that the
2QZ/6QZ can provide, errors in the estimate of the QSO OLF are largely
dominated by systematic effects (apart from possibly the very
brightest luminosities).  Therefore, although it is possible to
combine other surveys, in the analysis below, we chose not to include
QSOs from any other catalogue.  We also limit the analysis to regions
of the catalogue where the photometric completeness is greater than 85
per cent (see Fig. \ref{fig:colcomp}), corresponding to a redshift
range $0.4<z<2.1$.  In Fig. \ref{fig:zmb} we show the $z$ vs. $\mb$
distribution of sources, indicating the above cuts.

Within this redshift range, the corrections for photometric,
morphological and spectroscopic completenesses derived above were then
applied to this combined dataset.  To minimize the noise on the
estimate of the redshift-dependent spectroscopic incompleteness the
estimates presented in Table \ref{tab:fsz} were block-averaged over
$\Delta z = 0.3$ bins.

\subsection{$1/V$ estimator}

As in Paper I, we first obtain a graphical binned representation of
the OLF and its evolution with redshift using the $1/V$ estimator
devised by Page \& Carrera (2000).  We chose to bin the OLF into six
equally spaced linear redshift bins over the range $0.4<z<2.1$ and
into $\Delta \mb = 0.5$ bins in absolute magnitude.  All bins that
contain at least one QSO are plotted.  However, it should be noted
that bins near the magnitude limits of the survey can be biased even
using the Page \& Carrera estimator (see Paper I).  This is the case
if part of the bin lies outside of the survey limits, and there is
significant evolution across the bin.

The resulting OLF, calculated for a flat universe with $\om=0.3$,
$\ol=0.7$ and $\ho=70$\kmsmpc\ (which we call the $\Lambda$ cosmology)
is plotted in Fig. \ref{fig:olf}.  The value of the OLF, statistical
error and the number and mean redshift of QSOs contributing to each
plotted point in the figure is given in Table \ref{tab:olf}.  We
determine upper and lower 1$\sigma$ (84.13 per cent one-sided)
confidence intervals for a Poisson distribution \cite{gehrels86} for
bins which have small numbers of QSOs ($<20$), and hence have errors
which are not well represented by $\sqrt{N}$ errors.

\begin{table*}
\caption{Binned luminosity function estimate for $\om=0.3$, $\ol=0.7$
and $\ho=70$\kmsmpc, as plotted in Fig. \ref{fig:olf}.  We give the
value of $\lp$ in 6 redshift intervals, and in $\Delta\mb=0.5$ mag
bins.  We also list the mean redshift ($\bar{z}$) in each bin, the
number of QSOs contributing to the OLF ($\nq$) and the lower and upper
errors ($\dlp$).} 
\label{tab:olf}
\begin{tabular}{@{}ccrccccrccccrccc@{}}
\hline
&\multicolumn{5}{c}{$0.40<z<0.68$}&\multicolumn{5}{c}{$0.68<z<0.97$}&\multicolumn{5}{c}{$0.97<z<1.25$}\\
$\mb$ &$\bar{z}$&$\nq$&$\lp$&\multicolumn{2}{c}{$\dlp$}&$\bar{z}$ &$\nq$&$\lp$ &\multicolumn{2}{c}{$\dlp$}&$\bar{z}$ &$\nq$&$\lp$ &\multicolumn{2}{c}{$\dlp$} \\
\hline 
--20.75 & 0.43 &   31 & --5.62 & --0.09 & +0.07 &  --  &    0 &   --   &   --   &  --   &  --  &    0 &   --   &   --   &  --   \\
--21.25 & 0.47 &  166 & --5.65 & --0.04 & +0.03 &  --  &    0 &   --   &   --   &  --   &  --  &    0 &   --   &   --   &  --   \\
--21.75 & 0.54 &  393 & --5.67 & --0.02 & +0.02 & 0.69 &    2 & --5.46 & --0.45 & +0.36 &  --  &    0 &   --   &   --   &  --   \\
--22.25 & 0.57 &  484 & --5.73 & --0.02 & +0.02 & 0.73 &  184 & --5.70 & --0.03 & +0.03 &  --  &    0 &   --   &   --   &  --   \\
--22.75 & 0.57 &  329 & --5.92 & --0.02 & +0.02 & 0.80 &  553 & --5.79 & --0.02 & +0.02 & 0.99 &   18 & --5.68 & --0.12 & +0.11 \\
--23.25 & 0.58 &  256 & --6.04 & --0.03 & +0.03 & 0.84 &  710 & --5.83 & --0.02 & +0.02 & 1.06 &  469 & --5.81 & --0.02 & +0.02 \\
--23.75 & 0.58 &  139 & --6.28 & --0.04 & +0.04 & 0.84 &  513 & --5.99 & --0.02 & +0.02 & 1.12 &  863 & --5.86 & --0.02 & +0.01 \\
--24.25 & 0.60 &   61 & --6.56 & --0.06 & +0.05 & 0.86 &  340 & --6.18 & --0.02 & +0.02 & 1.12 &  754 & --5.94 & --0.02 & +0.02 \\
--24.75 & 0.62 &   18 & --7.00 & --0.12 & +0.11 & 0.87 &  155 & --6.50 & --0.04 & +0.03 & 1.13 &  452 & --6.18 & --0.02 & +0.02 \\
--25.25 & 0.61 &    5 & --7.54 & --0.24 & +0.22 & 0.88 &   62 & --6.80 & --0.06 & +0.05 & 1.13 &  225 & --6.49 & --0.03 & +0.03 \\
--25.75 & 0.63 &    5 & --7.48 & --0.24 & +0.22 & 0.86 &   18 & --7.22 & --0.12 & +0.11 & 1.14 &   83 & --6.87 & --0.05 & +0.05 \\
--26.25 & 0.63 &    1 & --7.98 & --0.77 & +0.52 & 0.80 &    4 & --7.87 & --0.28 & +0.25 & 1.10 &   23 & --7.25 & --0.10 & +0.08 \\
--26.75 &  --  &    0 &   --   &   --   &  --   & 0.86 &    3 & --7.98 & --0.34 & +0.29 & 1.16 &   10 & --7.59 & --0.16 & +0.15 \\
--27.25 &  --  &    0 &   --   &   --   &  --   &  --  &    0 &   --   &   --   &  --   & 1.17 &    2 & --8.29 & --0.45 & +0.36 \\

\hline
&\multicolumn{5}{c}{$1.25<z<1.53$}&\multicolumn{5}{c}{$1.53<z<1.81$}&\multicolumn{5}{c}{$1.81<z<2.10$}\\
$\mb$ &$\bar{z}$ &$\nq$&$\lp$ &\multicolumn{2}{c}{$\dlp$}&$\bar{z}$ &$\nq$&$\lp$ &\multicolumn{2}{c}{$\dlp$}&$\bar{z}$ &$\nq$&$\lp$ &\multicolumn{2}{c}{$\dlp$} \\
\hline
--23.75& 1.34 &  468 & --5.80 & --0.02 & +0.02 & 1.54 &    3 & --5.70 & --0.34 & +0.29 &  --  &    0 &   --   &   --   &  --  \\
--24.25& 1.40 & 1103 & --5.81 & --0.01 & +0.01 & 1.64 &  728 & --5.77 & --0.02 & +0.02 & 1.84 &   26 & --5.76 & --0.09 & +0.08\\
--24.75& 1.41 &  897 & --5.93 & --0.01 & +0.01 & 1.68 & 1142 & --5.85 & --0.01 & +0.01 & 1.93 &  897 & --5.85 & --0.01 & +0.01\\
--25.25& 1.41 &  495 & --6.20 & --0.02 & +0.02 & 1.68 &  823 & --6.02 & --0.02 & +0.01 & 1.96 &  898 & --6.00 & --0.01 & +0.01\\
--25.75& 1.41 &  228 & --6.55 & --0.03 & +0.03 & 1.68 &  470 & --6.27 & --0.02 & +0.02 & 1.95 &  618 & --6.18 & --0.02 & +0.02\\
--26.25& 1.42 &   95 & --6.89 & --0.05 & +0.04 & 1.69 &  224 & --6.60 & --0.03 & +0.03 & 1.96 &  333 & --6.45 & --0.02 & +0.02\\
--26.75& 1.44 &   24 & --7.30 & --0.10 & +0.08 & 1.71 &   82 & --6.98 & --0.05 & +0.05 & 1.97 &  120 & --6.90 & --0.04 & +0.04\\
--27.25& 1.31 &    2 & --8.36 & --0.45 & +0.36 & 1.72 &   16 & --7.50 & --0.12 & +0.12 & 1.98 &   43 & --7.24 & --0.07 & +0.06\\
--27.75& 1.48 &    2 & --8.36 & --0.45 & +0.36 & 1.71 &    5 & --8.00 & --0.24 & +0.22 & 2.07 &    2 & --8.42 & --0.45 & +0.36\\
--28.25&  --  &    0 &   --   &   --   &  --   &  --  &    0 &   --   &   --   &  --   & 1.94 &    2 & --8.42 & --0.45 & +0.36\\
\hline
\end{tabular}
\end{table*}

\subsection{Maximum likelihood analysis}

To obtain a more quantitative descriptions of suitable models we
carried out a maximum likelihood analysis, fitting a number of
models to the data (e.g. Boyle, Shanks \& Peterson 1988). This technique
relies on maximizing the likelihood function $S$ corresponding to the
Poisson probability distribution function for both model and data
\cite{mtaz83}. Previously, we have used the 2D K-S statistic to
provide a goodness-of-fit measure for `best fit'  maximum likelihood
models.  However, the data-set is now so large that the K-S test,
notoriously insensitive to discrepancies between the data and the
model predictions in the wings of the distributions, is very poor at
discriminating between models.  Instead, we choose to use the $\chi^2$
statistic, based on the numbers of observed and predicted QSOs in each
of the $\mb,z$ bins used for the $1/V$ estimator above.

In keeping with our general philosophy of previous maximum likelihood
fits to the OLF, we used the minimum numbers of free parameters in any
model required to obtain an acceptable fit.  Our $a\ priori$
definition of an acceptable fit was one which could not be  rejected
at the 99 per cent confidence level or greater based on the $\chi^2$
statistic.  Errors on the fit parameters correspond to the $\Delta S =
1$ contours around each parameter, or, equivalently, the 68 per cent
confidence contour for one interesting parameter.

While carrying out the maximum likelihood analysis, we employ an
absolute magnitude limit $\mb>-22.5$ for the $\Lambda$ cosmology and
$\mb>-22$ for an Einstein-de Sitter (EdS) cosmology ($\om=1.0$,
$\ol=0.0$ and $\ho=70$\kmsmpc).  This absolute magnitude cut excludes
AGN where the contribution from the host galaxy (either
photometrically or spectroscopically) may lead to a significantly bias
against the inclusion or identification of such objects in the 2QZ
which is not accurately modelled by the above completeness analysis.
The calculated absolute magnitude of these sources would also be
biased by the inclusion of flux from the host galaxy.

\begin{table*}
\caption{The best fit OLF model parameters from the maximum likelihood
analysis for the range of models discussed.  We list the redshift
range and faint $\mb$ limit of the data fitted, the evolution model and
cosmology used (assuming a flat $\om+\ol=1$ Universe), the number of
QSOs in the analysis ($\nq$) and the best fit values of the model
parameters.  We also give the $\chi^2$ value for the comparison of the
model to the data, as well as the number of degrees of freedom ($\nu$)
and the $\chi^2$ probability.}
\label{tab:params}
\begin{tabular}{@{}cccccccccccrrc@{}}
\hline
Redshift& $\mb$ & Evolution & $\om$&  $\nq$ & $\alpha$ & $\beta$ & $\mb^*$ & $k_{1}$ & $k_{2}$ &  $\Phi^{*}$ & $\chi^2$ & $\nu$ & $P_{\chi^2}$ \\
range& limit & model & & & & & & & &  ${\rm Mpc}^{-3}{\rm mag}^{-1}$ &\\
\hline 
$0.4<z<2.1$ & --22.5&$k_{1}z + k_{2}z^{2}$ &0.3 & 15830 & --3.31 & --1.09 & --21.61 &1.39 & --0.29 &1.67$\times 10^{-6}$&71.1  & 62 & 2.00e-01 \\ 
$0.4<z<2.1$ & --22  &$k_{1}z + k_{2}z^{2}$ &1.0 & 15875 & --3.28 & --1.12 & --20.74 &1.49 & --0.31 &5.07$\times 10^{-6}$&148.6 & 61 & 2.88e-09 \\ 
$0.4<z<2.1$ & --22.5&$e^{k\tau}$           &0.3 & 15830 & --3.25 & --1.01 & --20.47 &6.15 & --     &1.84$\times 10^{-6}$&64.5  & 63 & 4.24e-01 \\ 
$0.4<z<2.1$ & --22  &$e^{k\tau}$           &1.0 & 15875 & --3.16 & --0.98 & --18.54 &7.16 & --     &5.99$\times 10^{-6}$&148.6 & 62 & 2.52e-07 \\ 
\hline
\end{tabular}
\end{table*} 

\begin{table*}
\centering
\caption{The number of observed, $\nobs$, and predicted, $\npre$ QSOs
for the best fit exponential evolution model in $\Lambda$ Universe,
along with the significance of the devation away from the model,
$\sigma$.}
\label{tab:chi2}
\setlength{\tabcolsep}{3pt}
\begin{tabular}{@{}crrrrrrrrrrrrrrrrrr@{}}
\hline
&\multicolumn{3}{c}{$0.40<z<0.68$}
&\multicolumn{3}{c}{$0.68<z<0.97$}
&\multicolumn{3}{c}{$0.97<z<1.25$}
&\multicolumn{3}{c}{$1.25<z<1.53$}
&\multicolumn{3}{c}{$1.53<z<1.81$}
&\multicolumn{3}{c}{$1.81<z<2.10$}\\
$\mb$ & \multicolumn{1}{c}{$\nobs$} & \multicolumn{1}{c}{$\npre$} & \multicolumn{1}{c}{$\sigma$} &\multicolumn{1}{c}{$\nobs$} & \multicolumn{1}{c}{$\npre$} & \multicolumn{1}{c}{$\sigma$} &\multicolumn{1}{c}{$\nobs$} & \multicolumn{1}{c}{$\npre$} & \multicolumn{1}{c}{$\sigma$} &\multicolumn{1}{c}{$\nobs$} & \multicolumn{1}{c}{$\npre$} & \multicolumn{1}{c}{$\sigma$} &\multicolumn{1}{c}{$\nobs$} & \multicolumn{1}{c}{$\npre$} & \multicolumn{1}{c}{$\sigma$} &\multicolumn{1}{c}{$\nobs$} & \multicolumn{1}{c}{$\npre$} & \multicolumn{1}{c}{$\sigma$} \\
\hline 
--22.75 & 329&  354.13&  --1.34&  553&  563.35&  --0.44&   18&   12.47&    1.20&   --&     -- &     -- &   --&     -- &     -- &   --&     -- &     --  \\
--23.25 & 256&  254.05&    0.12&  710&  698.62&    0.43&  469&  480.41&  --0.52&    0&    0.42&  --1.01&   --&     -- &     -- &   --&     -- &     --  \\
--23.75 & 139&  134.55&    0.38&  513&  546.66&  --1.44&  863&  892.78&  --1.00&  468&  490.51&  --1.02&    3&    3.40&  --0.23&   --&     -- &     --  \\
--24.25 &  61&   51.30&    1.35&  340&  336.81&    0.17&  754&  731.82&    0.82& 1103& 1030.66&    2.25&  728&  722.35&    0.21&   26&   20.07&    1.32 \\
--24.75 &  18&   13.40&    0.97&  155&  160.39&  --0.43&  452&  467.05&  --0.70&  897&  821.93&    2.62& 1142& 1151.22&  --0.27&  897&  883.79&    0.44 \\
--25.25 &   5&    4.68&    0.10&   61&   55.87&    0.69&  225&  233.01&  --0.52&  495&  505.52&  --0.47&  823&  839.19&  --0.56&  898&  958.27&  --1.95 \\
--25.75 &   5&    1.61&    1.36&   18&   14.08&    0.81&   83&   90.69&  --0.81&  228&  243.56&  --1.00&  470&  469.90&    0.00&  618&  611.27&    0.27 \\
--26.25 &   1&    0.45&    0.27&    4&    4.98&  --0.46&   23&   22.21&    0.17&   95&   94.76&    0.02&  224&  210.91&    0.90&  333&  302.49&    1.75 \\
--26.75 &   0&    0.05&  --0.03&    3&    1.76&    0.49&   10&    7.27&    0.72&   24&   21.84&    0.46&   82&   75.41&    0.76&  120&  125.28&  --0.47 \\
--27.25 &  --&     -- &     -- &    0&    0.49&  --1.03&    2&    2.59&  --0.39&    2&    7.18&  --1.98&   16&   16.59&  --0.15&   43&   40.26&    0.43 \\
--27.75 &  --&     -- &     -- &    0&    0.04&  --0.02&    0&    0.88&  --1.16&    2&    2.56&  --0.37&    5&    5.74&  --0.32&    2&    9.49&  --2.48 \\
--28.25 &  --&     -- &     -- &   --&     -- &     -- &    0&    0.11&  --0.53&    0&    0.89&  --1.16&    0&    2.04&  --1.56&    2&    3.38&  --0.79 \\
--28.75 &  --&     -- &     -- &   --&     -- &     -- &   --&     -- &     -- &    0&    0.11&  --0.51&    0&    0.69&  --1.09&    0&    1.20&  --1.28 \\
--29.25 &  --&     -- &     -- &   --&     -- &     -- &   --&     -- &     -- &   --&     -- &     -- &    0&    0.06&  --0.08&    0&    0.38&  --1.00 \\
--29.75 &  --&     -- &     -- &   --&     -- &     -- &   --&     -- &     -- &   --&     -- &     -- &   --&     -- &     -- &    0&    0.01&    0.00 \\
\hline
\end{tabular}
\end{table*} 

\subsection{Model fitting}

\begin{figure}
\centering 
\centerline{\psfig{file=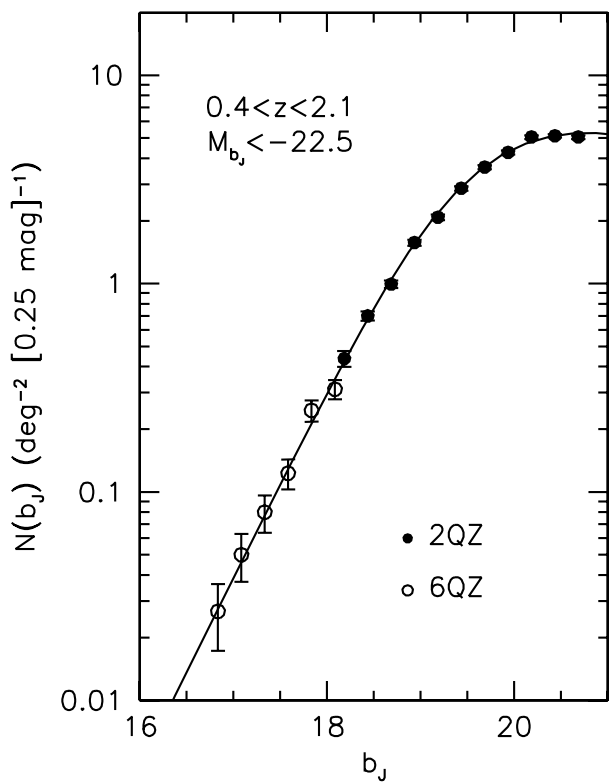,width=12cm}}
\vspace{-0.5truecm}
\caption{Differential number-magnitude, $n(\bj)$, relation for the 2QZ
(filled circles) and 6QZ (open circles) for QSOs with $0.4<z<2.1$ and
$\mb<-22.5$.  Also shown is the prediction from the best fit
exponential evolution model in the $\Lambda$ Universe.}
\label{fig:nm} 
\end{figure}

\begin{figure}
\centering 
\centerline{\psfig{file=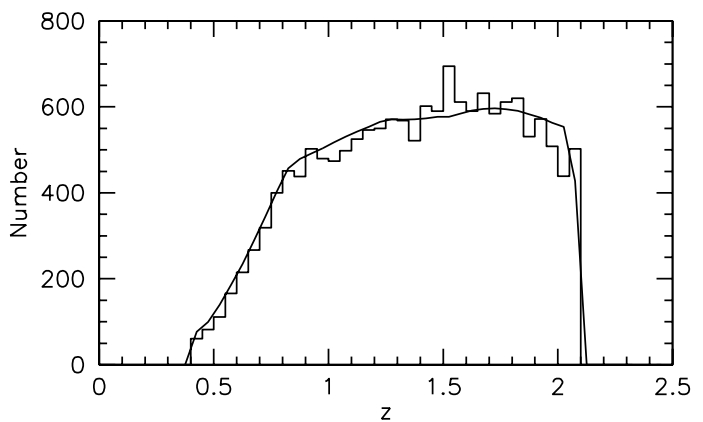,width=10cm}}
\vspace{-3.5truecm}
\caption{Observed number-redshift, $n(z)$, relation for the
2QZ survey for $0.4<z<2.1$ and $\mb<-22.5$.  The solid line
indicates the best fit exponential evolution model for the 
$\Lambda$ Universe.}
\label{fig:nz} 
\end{figure}

Based on our previous results with the 2QZ, we chose to model the OLF
$\Phi(L, z)$ as a double power law in luminosity such that
\begin{equation}
\Phi (\lb ,z) = \frac{\Phi(\lb^*)}{(\lb/\lb^*)^{-\alpha} 
+ (\lb/\lb^*)^{-\beta}}.
\end{equation}
Expressed in magnitudes this becomes
\begin{equation}
\Phi (\mb,z) = \frac{\Phi(\mb^{*})}{10^{0.4(\alpha+1)(\mb-
\mb^*)} + 10^{0.4(\beta+1)(\mb-\mb^*)}}.
\end{equation}
The evolution of the luminosity function is given by the redshift
dependence of the break luminosity $\lb^*\equiv\lb^*(z)$ or magnitude,
$\mb^*\equiv\mb^*(z)$.  In earlier papers in this series, we have
modelled this evolution as a 2nd-order polynomial of the form
\begin{equation}
\lb ^{*}(z) = \lb ^{*}(0)10^{k_1z+k_2z^2},
\end{equation}
or equivalently,
\begin{equation}
\mb^{*}(z) = \mb^{*}(0) - 2.5 (k_1 z + k_2z^2).
\end{equation}
In this paper we also attempt to fit an exponential 
evolution with look-back time ($\tau$) of the form
\begin{equation}
\lb ^{*}(z) = \lb ^{*}(0)\exp(k_1\tau).
\end{equation}
Expressing this in magnitudes it becomes
\begin{equation}
\mb^{*}(z) = \mb^{*}(0) - 1.08k_1\tau
\end{equation}
Although such models with an exponential evolution in $\tau$ have
previously been shown not to fit the QSO distribution
\cite{paper1,lc97} for an EdS Universe, exponential evolution models
have not been fitted to Universes with non-zero $\Omega_\Lambda$.

These models were fitted to the survey data for both EdS and $\Lambda$
cosmologies.  The statistical errors on individual parameters are:
$\sigma(\alpha) = 0.05$, $\sigma(\beta) = 0.1$, $\sigma(\mb^*) =
0.2$, $\sigma(k_1) ({\rm exponential}) = 0.1$, $\sigma(k_1) ({\rm
polynomial}) = 0.03$, $\sigma(k_2)  ({\rm polynomial}) = 0.01$.  We
note that the size of the 2QZ/6QZ sample is such that residual systematic
uncertainties are likely to dominate the model fitting.  The
completeness analysis described above (Section \ref{sec:comp}) is
aimed at minimizing such uncertainties.  Although not affecting the
shape of the fitted QSO OLF, any error in the survey $\bj$ zero-point
will translate to a shift in $\mb^*$.  Checks against a small number
of independent CCD sequences in the 2QZ region suggest that such an
error must be no greater than 0.05 mag.

\subsection{Results}

We find that both the polynomial evolution in $z$ and the exponential
evolution in $\tau$  provide acceptable fits when using the $\Lambda$
cosmology.  The best fit model values are shown in Table
\ref{tab:params}, along with their derived $\chi^2$ values and
probability of acceptance. The best fit exponential model using the
$\Lambda$ cosmology has a rate of evolution defined by $k_1=6.15$.
This corresponds to an 'e-folding' time of approximately 2Gyr.  Both
evolutionary models provide a poor fit to the full data set in an EdS
Universe.  Although these model fits are rejected, for completeness we
also show the best fit model parameters for an EdS Universe in Table
\ref{tab:params}.

The predicted differential number-magnitude, $n(\bj)$, and
number-redshift, $n(z)$, relations for the 2QZ and 6QZ surveys based
on the best fit exponential evolution model in a $\Lambda$ Universe
are shown in Fig. \ref{fig:nm} and Fig. \ref{fig:nz}.  The model
predictions provide a good fit to the derived $n(\bj)$ and observed
$n(z)$ relations for QSOs over the redshift range $0.4<z<2.1$ and with
absolute magnitudes $\mb<-22.5$. The model also provides a good match
to the binned OLF (see Fig. \ref{fig:olf}), although some small
differences can be seen.  In  Fig. \ref{fig:olf} the model points are
derived based on the accessible volume in each bin (as for the data
points).  This adjusts the model for biases introduced by binning in
exactly the same way as the data.  As a result, the model curves
(dotted lines) do not trace out perfect double power laws, although
they are based on the double power law model.

To investigate any systematic deviations of the data away from the
model we show the number of objects observed and predicted, and the
significance of the difference between the two, $\sigma=(N_{\rm
obs}-N_{\rm pred})/\sqrt{N_{\rm pred}})$ in Table \ref{tab:chi2} for
the exponential model in a $\Lambda$ Universe.  Positive values of
$\sigma$ correspond  to cases in which there are more QSOs observed
than predicted.  $\sqrt{N_{\rm pred}}$ is only a reasonable estimate
of the error for large $N_{\rm pred}$.  For small $N_{\rm pred}$
($<20$) we use the 84.13 per cent one-sided confidence interval in a
Poisson distribution to determine the error.  As the Poisson
distribution is not a symmetric function, we choose the error estimate
depending which side of the model expectation the observed number of
QSOs lies.  While not giving the formally correct significance for an
arbitrary number of standard devations (as the Poisson distribution
still has a different shape to a Gaussian distribution) this method
does produce a more realistic estimate of how significant the
deviations from the model are.

It is clear from Fig. \ref{fig:olf} and Table \ref{tab:chi2} that
although the model is in general good agreement with the data, there
are some deviations away from the model.  The most significant of
these are at relatively faint magnitudes near the break in the OLF
(e.g. at $\mb\sim-24.0$ in the $1.25<z<1.53$ interval).  Although the
deviations are small, these bins contain large numbers of QSOs, making
the errors on each bin small.  It is possible that the simple double
power law parameterization is not a perfect description of the data
when sufficient statistical precision is available.  It is also
possible that residual systematic errors are beginning to dominate the
fitting.  As a second goodness of fit test we also calculate the 2-D
K-S probability for the exponential model in the  $\Lambda$ cosmology,
and find a value of $6.4$ per cent, making this model acceptable at
the $2\sigma$ level and under our pre-defined criteria (models
should not be rejected at $>99$ per cent level).  At the bright end of
the OLF there is some  evidence of the steepening of the bright end
slope towards higher redshift found by other authors (e.g. Goldschmidt
\& Miller 1998), with the model slightly under-prediciting the number
of bright low redshift QSOs and similarly over-prediciting the numbers
of bright high redshift QSOs.  By simply examining the numbers
observed and predicited we find there is a $2\sigma$ excess of QSOs in
our lowest redshift bin at $\mb<-24.0$ (90 observed vs. 71.49
predicted), and a $1\sigma$ deficit of QSOs in our highest redshift
bin at $\mb<-27.0$ (47 observed vs. 54.72 predicited).  We note that
these deviations from the model are less significant than the those
found at fainter magnitudes, discussed above.

\begin{figure}
\centering 
\centerline{\psfig{file=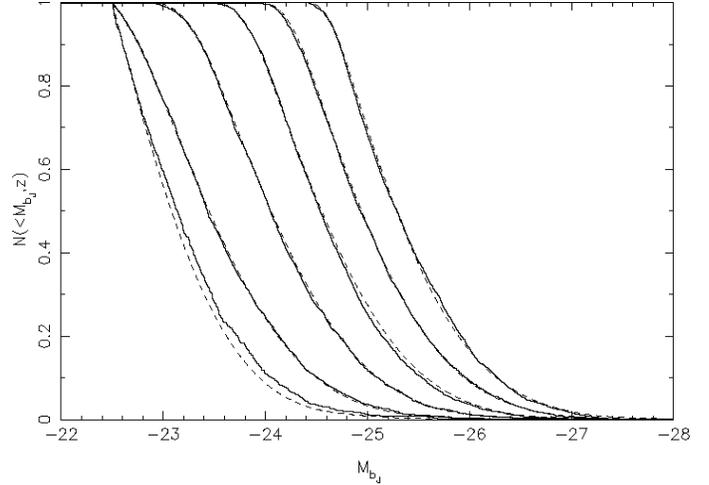,width=9cm,angle=270}}
\caption{The normalized cumulative distribution of QSOs in each of our
six redshift intervals (solid lines) going from low (left) to high
(right) redshift.  These are the raw observed numbers limited to
$\mb<-22.5$ in the $\Lambda$ cosmology.  Also plotted are the
predicted cumulative distributions from the best fit exponential
evolution model (dashed lines) after correcting them for
incompleteness.}
\label{fig:lf_ks1d} 
\end{figure}

To further examine deviations from the best fit model we carry out an
analysis similar to Goldschmidt \& Miller (1998).  We compare the
cumulative distribution of the number of QSOs observed and predicted
as a function of $\mb$ for our six separate redshift slices.  This
test has the advantage of not relying on comparisons to binned
luminosity functions.  Both the data (solid lines) and models (dashed
lines) are shown in Fig \ref{fig:lf_ks1d}.  We use the
best fit model using the exponential evolution in look-back time in a
$\Lambda$ cosmology.  Both the model and data are limited to
$\mb<-22.5$ in order to compare only the regions fitted.  We see that
there is good general agreement between the model and data.  However,
we do see that the data in the lowest redshift interval
($0.40<z<0.68$) contains more bright QSOs than predicted by the model.
We carry out a 1D K-S test comparing the model and data in each of our
six redshift slices.  The resulting probabilities of the null
hypothesis that the data are drawn from a distribution described by
the model are 0.037, 0.725, 0.708, 0.068, 0.806, 0.099 for the low to
high redshift intervals respectively.  The largest deviation is found
in the lowest redshift interval, in which the data deviate from the
model at the $\sim2\sigma$ level.  This agrees with the excess
found by examining the numbers of predicted and observed low redshift
QSOs at $\mb<-24$ and is further evidence of the excess of
bright low redshift QSOs found by other authors (e.g. Goldschmidt \&
Miller 1998).  If we examine the binned OLF in
Fig. \ref{fig:olf}, we see that the excess over the model appears to
occur at intermediate magnitudes, $\mb\sim-24.0$ and brighter.

We also test the effect of extending the range of our fits in both
absolute magnitude and redshift.  When fitting the exponential
evolution model in a $\Lambda$ cosmology we obtain an acceptable fit
with a faint limit of $\mb<-22$ rather than our adopted limit of
$\mb<-22.5$, with parameters virtually indistinguishable from the
best fit values shown in Table \ref{tab:params} for the $\mb<-22.5$
limit.  Extending the redshift range of our fits we find that
acceptable fits to the exponential evolution model in a $\Lambda$
cosmology are found in the redshift range $0.4<z<2.2$, but not at
higher redshift (e.g. $0.4<z<2.3$).  Similar results are found with
the polynomial evolution model, with an acceptable fit at $0.4<z<2.2$,
but not with $0.4<z<2.3$.  This is indicative of either (or both) a
change in the evolution of QSOs or (and) possible residual systematic
errors in the large completeness corrections required at high
redshift.  

This appears to demonstrate that a sample of the size of the 2QZ is
limited by the accuracy of corrections for systematic imcompleteness,
rather than random noise.  It is also very likely that although the
simple model fits above provide a good general description for the
evolution of the QSO population, the actual QSO luminosity function,
including its evolution, has a more complex form.  Evolution in the
QSO OLF appears to turn over, peaking at $z\sim2-3$, with declining
QSO activity at higher redshift (e.g. Warren, Hewett \& Osmer 1994;
Fan et al. 2001).  There are a number of further analyses that could
be carried out on the QSO OLF, however we leave these to future papers.  

The physical mechanisms which drive the shape and evolution of the QSO
OLF are still uncertain.  The two power law model fits the data well,
but the observed break may be due to a number of phenomena.  The break
in the OLF could correspond to a break in the black hole mass
function, which in turn is related to the mass (or velocity
dispersion) of the host galaxy.  Black hole mass is seen to correlate
well with the velocity dispersion of the spheroidal component of local
galaxies (e.g. Gebhardt et al 2000), although this has yet to be
observed at high redshift.  However, the correlation between host
galaxy and AGN {\it luminosity} is less clear with only a weak
correlation with large scatter found between the two (e.g. Schade et
al. 2000), at least at low redshift.  This implies that AGN have a
broad range of efficiencies, with some accreting near (or above) the
Eddington limit and others accreting at a much lower rate.  Such a
broad range in accretion rates convolved with any black hole mass
function would contrive to significantly blur any break.  However, it
is likely, although by no means certain, that only a small fraction of
sources accrete at super-Eddington rates.  This then provides a second
natural cut off at high luminosity.  In this context we also need to
understand the cause of the strong luminosity evolution seen in the
QSO population.  One view is that this is driven by a decline in
accretion rate towards low redshift, as the available fuel becomes
more scarce.  A second factor, is that possible tiggering events
(e.g. mergers) are more common at high redshift.  The current simple
semi-analytic models of QSO formation (e.g. Kauffmann \& Haehnelt
2000) predict that black hole masses will be lower at high redshift,
but that these AGN will be accreting closer to the Eddington limit.
This implies that for a given luminosity, black hole masses will be
lower at high redshift.  This model is called into question by the
observation of no correlation between QSO line widths (which should
provide an estimate of black hole mass) and redshift
\cite{composite03}.  It is clear that the QSO OLF, along with other
observations of black hole mass estimates and host galaxy properties
have the ability to help us understand the underlying physics driving
AGN formation and evolution.

The 2QZ and 6QZ surveys are being made available to the astronomical
community.  It is hoped that they will provide a valuable and lasting
resource for astronomers, and allow greater insight into the physical
processes governing AGN as well as enhancing our understanding of
cosmology and the structure of the Universe at high redshift.

\section*{ACKNOWLEDGMENTS} 

We warmly thank all the present and former staff of the
Anglo-Australian Observatory for their work in building and operating
the 2dF and 6dF facilities.  The 2QZ and 6QZ are based on observations
made with the Anglo-Australian Telescope and the UK Schmidt Telescope.
We also indebted to Mike Irwin and all the staff at the APM facility
for the scanning of the UKST plates that are the basis of the 2QZ and
6QZ surveys.  We thank Mike Hawkins for providing us with colour
information on his QSO sample prior to publication.  We also thank
Mike Brotherton for allowing us to include the IDs based on the Keck
observations of NVSS sources in the 2QZ.  Finally we wish to thank the
referee, Paul Hewett, for his help in improving the final version of
this paper.

\vspace{1.0truecm}
This paper has been produced using the Blackwell Scientific Publications 
\TeX macros.

\end{document}